\documentclass[aps,prd,twocolumn,showpacs,showkeys,superscriptaddress,preprintnumbers,floatfix,reprint]{revtex4}
\usepackage{graphicx}
\usepackage{dcolumn}
\usepackage{amsmath}
\usepackage{amsfonts}
\usepackage{bm}
\usepackage{yhmath}
\usepackage{hyperref}
\usepackage{subfigure}
\usepackage{color}
\usepackage{xcolor}

\usepackage{multirow}
\usepackage{dcolumn}
\newcolumntype{C}[1]{>{\centering\arraybackslash}p{#1}}

\begin{document}

\preprint{
\vbox{
\hbox{ADP-16-27/T982}
}}

\def\Tr{{\rm Tr}}
\newcommand{\sla}{\not\!}
\def\sca{0.23}

\newenvironment{figure**}[1][]{\begin{widetext}
\begin{figure}[#1]
\begin{minipage}{\textwidth}
}{\end{minipage}
\end{figure}
\end{widetext} }

\allowdisplaybreaks

\title{Hamiltonian effective field theory study of the $\mathbf{N^*(1440)}$ resonance in lattice QCD}

\author{Zhan-Wei Liu}\email{zhan-wei.liu@adelaide.edu.au}
\affiliation{Centre for the Subatomic Structure of Matter (CSSM), Department of Physics, 
             University of Adelaide, Adelaide SA 5005, Australia}

\author{Waseem Kamleh}
\affiliation{Centre for the Subatomic Structure of Matter (CSSM), Department of Physics, 
             University of Adelaide, Adelaide SA 5005, Australia}

\author{Derek B. Leinweber}
\affiliation{Centre for the Subatomic Structure of Matter (CSSM), Department of Physics, 
             University of Adelaide, Adelaide SA 5005, Australia}

\author{Finn M. Stokes}
\affiliation{Centre for the Subatomic Structure of Matter (CSSM), Department of Physics, 
             University of Adelaide, Adelaide SA 5005, Australia}

\author{Anthony W. Thomas}
\affiliation{Centre for the Subatomic Structure of Matter (CSSM), Department of Physics, 
             University of Adelaide, Adelaide SA 5005, Australia}
\affiliation{ARC Centre of Excellence in Particle Physics at the Terascale, Department of Physics, 
             University of Adelaide, Adelaide SA 5005, Australia}

\author{Jia-Jun Wu}
\affiliation{Centre for the Subatomic Structure of Matter (CSSM), Department of Physics, 
             University of Adelaide, Adelaide SA 5005, Australia}

\begin{abstract}

We examine the phase shifts and inelasticities associated with the $N^*(1440)$ Roper resonance and
connect these infinite-volume observables to the finite-volume spectrum of lattice QCD using
Hamiltonian effective field theory.
We explore three hypotheses for the structure of the Roper resonance.
All three hypotheses are able to describe the scattering data well.
In the third hypothesis the Roper resonance couples the low-lying bare basis-state component associated with the ground state nucleon with the virtual meson-baryon contributions.
Here the non-trivial superpositions of the meson-baryon scattering states are complemented by bare basis-state components explaining their observation in contemporary lattice QCD calculations.
The merit of this scenario lies in its ability to not only describe the observed nucleon energy levels in large-volume lattice QCD simulations but also explain why other low-lying states have been missed in today's lattice QCD results for the nucleon spectrum. 

\end{abstract}



\maketitle

\pagenumbering{arabic}

\section{Introduction}\label{secIntr}

Understanding the nature and structure of the excited states of the
nucleon is a key contemporary problem in QCD.  The Roper resonance,
$N^*(1440)$, was first deduced from the analysis of $\pi N$ phase shifts
in 1963 \cite{Roper1964}. However, its structure and nature have
aroused interest ever since \cite{Aznauryan2008,Joo2005}; it is lighter than the first odd-parity
nucleon excitation, $N^*(1535)$, and has a significant branching ratio
into $N\pi\pi$.  Although it is recognized as a well established
resonance (four-star ranking in the Review of Particle Physics)
\cite{PDG2014}, the properties of the Roper, such as the mass, width,
and decay branching ratios, still suffer large experimental
uncertainties \cite{Mokeev2012,Aznauryan2012,Thiel2015}.

On the theoretical side, there are widely varying models describing
the Roper resonance, such as early classical quark models
\cite{Weber1990,Julia-Diaz2006,Barquilla-Cano2007,Golli:2007sa,Golli:2009uk}, bag
\cite{Meissner1984} and Skyrme models \cite{Hajduk1984}, dynamically
generated by meson-nucleon interactions \cite{Krehl2000,
  Schuetz1998,Matsuyama2007,Kamano2010,Kamano2013,Hernandez2002a}, or
a monopole gluonic excitation
\cite{Barnes1983,Golowich1983,Kisslinger1995}.  However, these
descriptions do encounter challenges.  For example, predictions of the
mass are often too large or predictions for its width are too small.
Difficulties are also encountered in explaining its electromagnetic
coupling \cite{Sarantsev2008}.

One expects that lattice QCD simulations will provide unique
information concerning the Roper in a finite volume
\cite{Mahbub:2013ala,Mahbub:2010rm,Roberts:2013ipa,Roberts:2013oea,Alexandrou2015,Edwards2011,Kiratidis2015,Liu2014a}.
Current simulation results near the physical quark masses on lattices
with spatial length $L \simeq 3$ fm
\cite{Mahbub:2013ala,Mahbub:2010rm,Alexandrou2015,Kiratidis2015} reveal a $2S$-like
radial excitation \cite{Roberts:2013ipa,Roberts:2013oea} of the
nucleon near 1800 MeV, much higher than the infinite volume mass of
1440 MeV.  The main task of this paper is to examine the physical
phase shifts, inelasticities and pole position associated with the
$N^*(1440)$ Roper resonance and connect these infinite-volume
observables to the finite-volume spectrum of lattice QCD.  We use the
formalism of Hamiltonian effective field theory (HEFT) to achieve this
goal and seek an understanding of the observed finite-volume spectrum
in the context of empirical scattering observables.

The investigations of recent papers \cite{Hall2013, Wu2014} have shown
how to relate the eigenvalues of a finite-volume Hamiltonian matrix to
the spectrum of states observed in lattice QCD.  These two papers
explored the spectrum of states with the quantum numbers of the
$\Delta(1232)$ resonance \cite{Hall2013} and the $\pi\pi$-$K\bar{K}$
system \cite{Wu2014} via solutions of the eigen-equation of a
finite-volume Hamiltonian matrix.  Both papers showed that this
Hamiltonian matrix approach is equivalent to the well known L\"uscher
formulation \cite{Luscher1990,Luescher1991}.  Furthermore,
Ref.~\cite{Wu2014} showed that this method is sufficient for the
multi-channel scattering case, where the L\"uscher method is more
difficult to apply because it needs the phase shifts and inelastic
factors in every channel.  In contrast, the parameters of the
Hamiltonian can be constrained by the empirical phase shifts and
inelasticities.  As a result, the spectrum is easily obtained in HEFT.

This work is a direct application of HEFT in the $N\frac{1}{2}^+$
multi-channel case including three channels, $\pi N$, $\pi\Delta$, and
$\sigma N$. The parameters of the Hamiltonian are fitted to describe the phase shifts and inelasticities of $\pi N$
scattering up to 1800 MeV. Then, in the finite-volume relevant to
lattice QCD, the energy eigenvalues and their associated eigenvectors (describing the wave functions of the eigenstates) are obtained from the
Hamiltonian matrix.  Both the energy eigenvalues and the eigenstate
wave functions are important in understanding the spectrum of the
Roper channel obtained in today's lattice QCD simulations.

The framework of HEFT is described in Sec.~\ref{secFrm}.  We illustrate how
 the phase shifts and inelasticities in the infinite volume of
Nature are obtained and the manner in which the finite-volume energy
eigenstates are calculated.  The numerical results and associated
discussion are presented in Sec.~\ref{secNum}.  Here we present
results for
three different hypotheses for the internal structure of the Roper.

In the first case, the Roper is postulated to have a triquark-like
bare or core component with a mass exceeding the resonance mass.  This
component mixes with attractive virtual meson-baryon contributions,
including the $\pi N$, $\pi \Delta$, and $\sigma N$ channels, to
reproduce the observed pole position.
In the second hypothesis, the Roper resonance is dynamically generated
purely from the $\pi N$, $\pi \Delta$, and $\sigma N$ channels.
In the third hypothesis, the Roper resonance is coupled to the
low-lying bare component associated with the ground state nucleon.
Through coupling with the virtual meson-baryon contributions the
scattering data and pole position are reproduced.  The merit of this
third approach lies in its ability to not only describe the observed
nucleon energy levels in large-volume lattice QCD simulations but also
explain why other low-lying states have been missed in today's lattice
QCD results.  Finally, a short summary is given in Sec.~\ref{secSum}.

\section{Framework}\label{secFrm}

In this section we provide a short introduction to HEFT and illustrate
how it is used in both infinite and finite volumes.  The Hamiltonian
interactions associated with the $N\frac{1}{2}^+$ resonance channel
are described in Sec.~\ref{secInt}.  In Sec.~\ref{secPha}, the
phase shifts and inelasticities are derived from the Hamiltonian model
and the pole positions of states are easily obtained via the
$T$-matrix.  The Hamiltonian is then momentum discretised for the
finite-volume of the lattice in Sec.~\ref{secHam} and the spectrum of
energy eigenstates is obtained by solving the Hamiltonian
eigen equation.

\subsection{Hamiltonian in Channels with $\mathbf{I(J^P)=\frac12(\frac12^+)}$} \label{secInt}

The main channels strongly coupled to the Roper are the $\pi N$, $\pi
\Delta$ and $\sigma N$ channels. In the rest frame, the Hamiltonian of
the system with $I(J^P)=\frac12(\frac12^+)$ has the following
energy-independent form
\cite{Matsuyama2007,Kamano2010,Kamano2013,Thomas1984},
\begin{eqnarray}
H = H_0 + H_I.
\label{eq:h}
\end{eqnarray}

The non-interacting part is 
\begin{eqnarray}
H_0 &=&\sum_{B_0} |B_0\rangle \, m^{0}_{B} \, \langle B_0|+ \sum_{\alpha}\int d^3\vec{k}\nonumber\\
&&  |\alpha(\vec{k})\rangle \, \left [ \sqrt{m_{\alpha_1}^2 + \vec{k}^2} +
\sqrt{m_{\alpha_2}^2 + \vec{k}^2}\, \right ] \langle\alpha(\vec{k})|,
\label{eq:h0}
\end{eqnarray}
where $B_0$ is the bare baryon (including a bare nucleon $N_0$ or a
bare Roper $R_0$) with mass $ m^{0}_{B}$ and $\alpha$ denotes the
included channels $\pi N$, $\pi \Delta$, and $\sigma N$. The masses
$m_{\alpha_1}$ and $m_{\alpha_2}$ are the masses of the meson and
baryon in the channel $\alpha$, respectively.

The interaction Hamiltonian of this system includes two parts
\begin{eqnarray}
H_I = g + v,\label{eq:hi}
\end{eqnarray}
where $g$ describes the vertex interaction between the bare particle
and the two-particle channels $\alpha$ 
\begin{eqnarray}
g &=& \sum_{\alpha\, B_0} \int d^3\vec{k} \left \{  \,
|\alpha(\vec{k})\rangle \, G^\dagger_{\alpha, B_0}(k)\, \langle B_0|
\right . \nonumber \\
&&\qquad\qquad\quad \left . + |B_0\rangle\, G_{\alpha, B_0}(k)\, \langle \alpha(\vec{k})| 
\, \right \} \, ,
\label{eq:int-g}
\end{eqnarray}
while the direct two-to-two particle interaction is defined by
\begin{eqnarray}
v = \sum_{\alpha,\beta} \int d^3\vec{k}\; d^3\vec{k}'\,
|\alpha(\vec{k})\rangle\,  V^{S}_{\alpha,\beta}(k,k')\, \langle
\beta(\vec{k}')| \, .
\label{eq:int-v}
\end{eqnarray}
For the vertex interaction between the bare baryon and two-particle
channels, the following form is used: 
\begin{eqnarray}
G_{\pi N, B_0}^2(k)&=&\frac{3 g_{B_0\pi N}^2 }{4\pi^2 f^2} \frac{k^2 u_{\pi N}^2(k)}{\omega_\pi(k)},\\
G_{\pi \Delta, B_0}^2(k)&=&\frac{g_{B_0\pi\Delta}^2 }{3\pi^2 f^2} \frac{k^2 u_{\pi \Delta}^2(k)}{\omega_\pi(k)},\\
G_{\sigma N, B_0}^2(k)&=&\frac{g_{B_0\sigma N}^2 }{4\pi^2} \frac{u_{\sigma N}^2(k)}{\omega_\sigma(k)},
\end{eqnarray}
where $f=92.4$ MeV is the pion decay constant, $\omega_X(k)=\sqrt{k^2+m_X^2}$ is the corresponding
energy, and $u_\alpha(k)$ is the regulator \cite{Leinweber:2003dg,Wang:2007iw}.  We consider the exponential form
\begin{equation}
u_\alpha(k)=\exp\left ( -\frac{k^2}{\Lambda_\alpha^2} \right ) \, ,
\end{equation}
where $\Lambda_\alpha$ is the regularization scale.
%
Although we adopt the exponential form, our main conclusions are not affected if other form factors
are used.  We have explicitly checked the selection of a dipole form factor
$u_\alpha(k)=\left(1+{k^2}/{\Lambda_\alpha^2}\right)^{-2}$.  The phase shifts and inelasticities
are fit well and we obtain similar finite-volume results for the three scenarios 
considered in Sec.~\ref{secNum}.

For the two-to-two particle interaction, we introduce the separable potentials for the following five channels
\begin{eqnarray}
V_{\pi N, \pi N}^S(k,k')&=&g^S_{\pi N}\frac{\bar G_{\pi N}(k)}{\sqrt{\omega_\pi(k)}} 
      \frac{\bar G_{\pi N}(k')}{\sqrt{\omega_\pi(k')}}\, ,\label{eqVspiN}\\
V_{\pi \Delta, \pi \Delta}^S(k,k')&=&g^S_{\pi \Delta}\frac{\bar G_{\pi \Delta}(k)}{\sqrt{\omega_\pi(k)}} 
      \frac{\bar G_{\pi \Delta}(k')}{\sqrt{\omega_\pi(k')}}\, , \label{eqVsPiDelta}  \\
V_{\pi N, \pi \Delta}^S(k,k')&=&g^S_{\pi N,\pi \Delta}\frac{\bar G_{\pi N}(k)}{\sqrt{\omega_\pi(k)}} 
      \frac{\bar G_{\pi \Delta}(k')}{\sqrt{\omega_\pi(k')}}\, , \label{eqVPiNPiDel} \\
V_{\sigma N, \sigma N}^S(k,k')&=&g^S_{\sigma N,\sigma N}\frac{\bar G_{\sigma N}(k)}{\sqrt{\omega_\sigma(k)}} 
      \frac{\bar G_{\sigma N}(k')}{\sqrt{\omega_\sigma(k')}}\, , \label{eqVSigmaN}\\
V_{\pi N, \sigma N}^S(k,k')&=&g^S_{\pi N,\sigma N}\frac{\bar G_{\pi N}(k)}{\sqrt{\omega_\pi(k)}} 
      \frac{\bar G_{\sigma N}(k')}{\sqrt{\omega_\sigma(k')}}\, , \label{eqVPiNSigmaN}
\end{eqnarray}
where $\bar G_{\alpha}(k)=G_{\alpha, B_0}(k)/g_{B_0 \alpha}$.

\subsection{Phase Shift and Inelasticity}\label{secPha}

The T-matrices for two particle scattering can be obtained by solving
a three-dimensional reduction of the coupled-channel Bethe-Salpeter
equations for each partial wave 
\begin{eqnarray}
T_{\alpha, \beta}(k,k';E)&=&V_{\alpha, \beta}(k,k';E)+\sum_\gamma \int
q^2 dq \times \\
&&\hspace{-1.0cm} V_{\alpha, \gamma}(k,q;E)\,
\frac{1}{E-\omega_\gamma(q)+i \epsilon}\,  T_{\gamma, \beta}(q,k';E)
\, , \nonumber
\end{eqnarray}
where $\omega_\alpha(k)$ is the center-of-mass energy of channel $\alpha$
\begin{equation}
\omega_{\alpha}(k)=\sqrt{m_{\alpha_1}^2+k^2}+\sqrt{m_{\alpha_2}^2+k^2},
\end{equation}
and the coupled-channel potential can be calculated from the interaction Hamiltonian
\begin{eqnarray}
V_{\alpha, \beta}(k,k') &=& \sum_{B_0}\, G^\dag_{\alpha, B_0}(k)\,
\frac{1}{E-m_B^0}\,  G_{\beta, B_0}(k') \nonumber \\
&&+V^S_{\alpha,\beta}(k,k') \, .
\label{eq:lseq-2}
\end{eqnarray}
With the normalization $\langle
\alpha(\vec{k})\, |\, \beta(\vec{k}^{\,\,'})\rangle =
\delta_{\alpha,\beta}\, \delta (\vec{k}-\vec{k}^{\,\,'})$, the S-matrix
for $\pi N \to \pi N$ is related to the T-matrix by
\begin{eqnarray}
S_{\pi N}(E_{\rm cm})&=&1 -2i\pi \frac{\omega_\pi(k_{\rm cm})\, \omega_N(k_{\rm cm})}{E_{\rm cm}}\, k_{\rm cm}\nonumber\\
&&\quad\quad  \times T_{\pi N,\pi N}(k_{\rm cm}, k_{\rm cm}; E_{\rm
  cm}) \, ,
\end{eqnarray}
where $k_{\rm cm}$ satisfies the on-shell condition $\sqrt{m_N^2+k_{\rm
    cm}^2}+\sqrt{m_\pi^2+k_{\rm cm}^2}=E_{\rm cm}$. One can obtain
phase shifts $\delta$ and inelasticities $\eta$ with $S_{\pi N}(E_{\rm
  cm})=\eta \exp(2i\delta)$.

In addition to the phase shifts and inelasticities, the pole positions
of bound states or resonances can also be obtained by searching for
the poles of the T-matrix.\\

\subsection{Finite-Volume Matrix Hamiltonian Model} \label{secHam}

We present the formalism of the finite-volume matrix Hamiltonian model by
following Refs.~\cite{Hall2013,Hall2015}. The Hamiltonian $\mathcal H$
at finite volume is the momentum discretisation of the Hamiltonian $H$ at
infinite volume. It can also be written as a sum of free and
interacting Hamiltonians $\mathcal H=\mathcal H_0+\mathcal H_I$.

In the center-of-mass frame, the meson and the baryon in the
two-particle states carry the same magnitude of momentum with
back-to-back orientation, while the bare baryon is at rest. In the
finite periodic volume of the lattice with length $L$, the momentum of
a particle is restricted to $k_n=2\pi\sqrt{n}/L$, with $n = n_x^2 +
n_y^2 + n_z^2$ such that $n=0,1,2,\ldots$

In the finite volume, it is convenient to express the Hamiltonian $\mathcal H$ as a matrix. $\mathcal H_0$ is a diagonal matrix
\begin{equation}
\mathcal H_0={\rm diag}\{m_B^0,\, \omega_{\sigma N}(k_0),~ \omega_{\pi
  N}(k_1),\, \omega_{\pi\Delta}(k_1),\, \omega_{\sigma N}(k_1),~...\}.
\end{equation}
The corresponding symmetric matrix $\mathcal H_I$ is
\begin{widetext}
\begin{equation}
\mathcal H_I=\left( \begin{array}{ccccccc}
                         0 & \mathcal  G_{\sigma N, B_0}(k_0) &                  \mathcal  G_{\pi N, B_0}(k_1) &                  \mathcal  G_{\pi \Delta, B_0}(k_1) & \mathcal  G_{\sigma N, B_0}(k_1) &                   \mathcal  G_{\pi N, B_0}(k_2) & \ldots \\
  \mathcal  G_{\sigma N, B_0}(k_0) &\mathcal  V^S_{\sigma N,\sigma N}(k_0,k_0) & \mathcal  V^S_{\sigma N,\pi N}(k_0,k_1)&                                           \mathcal  V^S_{\sigma N,\pi \Delta}(k_0,k_1) &  \mathcal  V^S_{\sigma N,\sigma N}(k_0,k_1)&   \mathcal  V^S_{\sigma N,\pi N}(k_0,k_2) & \ldots \\
     \mathcal  G_{\pi N, B_0}(k_1) &\mathcal  V^S_{\pi N,\sigma N}(k_1,k_0) &      \mathcal  V^S_{\pi N,\pi N}(k_1,k_1) &      \mathcal  V^S_{\pi N,\pi \Delta}(k_1,k_1) &  \mathcal  V^S_{\pi N,\sigma N}(k_1,k_1) &       \mathcal  V^S_{\pi N,\pi N}(k_1,k_2) & \ldots \\
\mathcal  G_{\pi \Delta, B_0}(k_1) & \mathcal  V^S_{\pi \Delta,\sigma N}(k_1,k_0) & \mathcal  V^S_{\pi \Delta,\pi N}(k_1,k_1) & \mathcal  V^S_{\pi \Delta,\pi \Delta}(k_1,k_1) &  \mathcal  V^S_{\pi \Delta,\sigma N}(k_1,k_1) & \mathcal  V^S_{\pi N, \pi \Delta}(k_1,k_2) & \ldots \\
  \mathcal  G_{\sigma N, B_0}(k_1) &\mathcal  V^S_{\sigma N,\sigma N}(k_1,k_0) & \mathcal  V^S_{\sigma N,\pi N}(k_1,k_1)&                                           \mathcal  V^S_{\sigma N,\pi \Delta}(k_1,k_1) &  \mathcal  V^S_{\sigma N,\sigma N}(k_1,k_1)&   \mathcal  V^S_{\sigma N,\pi N}(k_1,k_2) & \ldots \\
     \mathcal  G_{\pi N, B_0}(k_2) &\mathcal  V^S_{\pi N,\sigma N}(k_2,k_0) &      \mathcal  V^S_{\pi N,\pi N}(k_2,k_1) &     \mathcal  V^S_{\pi N, \pi \Delta}(k_2,k_1) &  \mathcal  V^S_{\pi N,\sigma N}(k_2,k_1) &       \mathcal  V^S_{\pi N,\pi N}(k_2,k_2) & \ldots \\
                    \vdots &                   \vdots &                                 \vdots &                                      \vdots &                         \vdots &                                  \vdots & \ddots \\
\end{array}
\right),
\end{equation}
\end{widetext}
where 
\begin{equation}
\mathcal G_{\alpha,B_0}(k_n)=\sqrt{\frac{C_3(n)}{4\pi}}\, \left
  (\frac{2\pi}{L}\right )^{3/2}\, G_{\alpha,B_0}(k_n)\, ,
\end{equation}
\begin{equation}
\mathcal V^S_{\alpha,\beta}(k_n,
  k_m)=\frac{\sqrt{C_3(n)\,C_3(m)}}{4\pi}\, \left
  (\frac{2\pi}{L}\right )^{3}\, 
  V^S_{\alpha,\beta}(k_n, k_m) \, .
\end{equation}
$C_3(n)$ represents the degeneracy factor for summing the squares of
three integers to equal $n$.

One can obtain the eigenstate energy levels on the lattice and analyse the
corresponding eigenvector wave functions describing the constituents
of the eigenstates with the above Hamiltonian $\mathcal H$. 

In addition to the results at physical pion mass, we can also extend
the formalism to unphysical pion masses.  Using $m_\pi^2$ as a measure
of the light quark masses, we consider the variation of the bare mass
and $\sigma$-meson mass as
\begin{eqnarray}
m_B^0(m_\pi^2)&=&m_B^0|_{\rm phy}+\alpha_B^0\, (m_\pi^2-m_\pi^2|_{\rm phy})\, , \\
m_\sigma^2(m_\pi^2)&=&m_\sigma^2|_{\rm phy}+\alpha_\sigma^0\, (m_\pi^2-m_\pi^2|_{\rm phy}),
\end{eqnarray}
where the slope parameter $\alpha_B^0$ is constrained by lattice QCD data from the CSSM.  In the
large quark mass regime where constituent quark degrees of freedom become relevant one expects
$m_\sigma \sim m_\sigma|_{\rm phy} + (2/3)\, \alpha_N^0 \, (m_\pi^2-m_\pi^2|_{\rm phy})$
\cite{Cloet:2002eg} providing $\alpha_\sigma^0 \simeq \frac43 \, m_\sigma|_{\rm phy} \,
\alpha_N^0$.  The nucleon and Delta masses away from the physical point, $m_N(m_\pi^2)$ and
$m_\Delta(m_\pi^2)$, are obtained via linear interpolation between the corresponding data of
lattice QCD.  With $\alpha_N^0 = 1.00~{\rm GeV}^{-1}$, $\alpha_\sigma^0 \simeq 0.67$.  In the
following results, we find the $\sigma N$ channel couples weakly and therefore our conclusions are
not sensitive to this value.

For the other parameters constrained by experimental data, it is difficult to predict their
quark-mass dependence. However, Refs.~\cite{Liu:2015ktc,Hall2013} show examples where lattice data
can be described well without a quark-mass dependence for the couplings.  A similar approach has
been employed successfully in chiral effective field theory, where one expands in small momenta and
masses about the chiral limit.  Using fixed couplings, lattice data is described over a wide range
of pion masses. In any event, the lightest pion mass considered herein is very close to the
physical pion mass. The couplings should not change significantly over this small change in pion
mass.

\section{Numerical Results and Discussion} \label{secNum}

\subsection{Fitting the Phase Shift and Inelasticity} \label{secFitData}

Here we examine the phase shifts and inelasticities associated with the $N^*(1440)$ Roper resonance
and connect these infinite-volume observables to the finite-volume spectrum of lattice QCD using
HEFT.  It is natural to think that the Roper resonance might be dominated by a bare state dressed
by meson-baryon states, like the nucleon, but some authors also propose that the Roper may be a
dynamically-generated molecular state arising purely from multi-particle meson-baryon interactions.
These considerations lead us to explore three hypotheses for the structure of the Roper resonance.

\begin{figure}[tbp]
\begin{center}
\subfigure{\includegraphics[width=1.0\columnwidth]{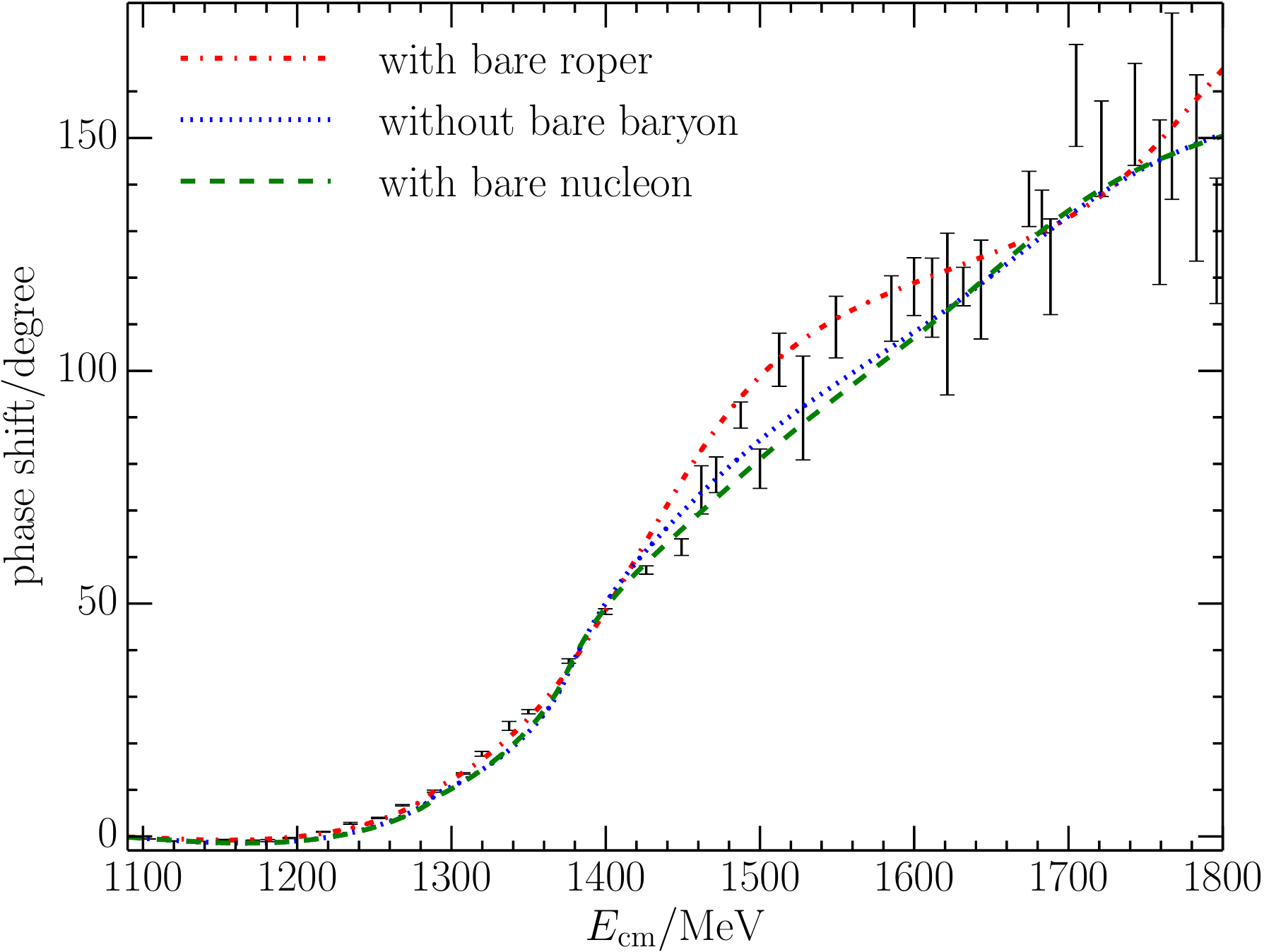}}
\subfigure{\includegraphics[width=1.0\columnwidth]{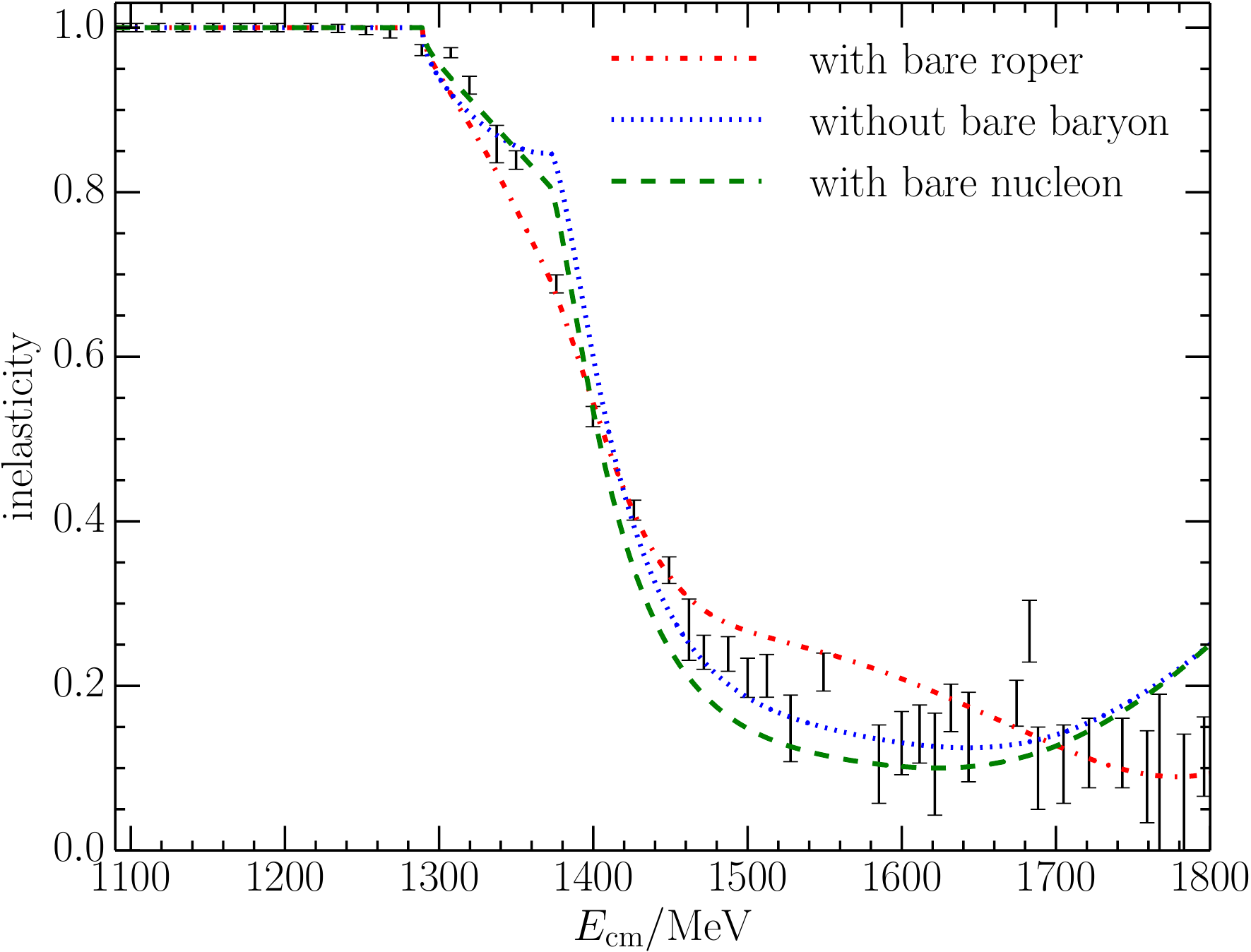}}
\caption{{\bf Colour online:} Phase shifts (upper) and inelasticities (lower) for $\pi N$ scattering
  with $I(J^P)=\frac12(\frac12^+)$.  The dot-dashed, dotted and dashed lines represent our best
  fits for scenario I with the bare Roper, scenario II without a bare baryon, and scenario III with
  the bare nucleon, respectively.  }\label{figPSEta}
\end{center} 
\end{figure}

In the first case, the Roper is postulated to have a triquark-like bare or core component with a
mass exceeding the resonance mass.  This component mixes with virtual meson-baryon
contributions, including the $\pi N$, $\pi \Delta$, and $\sigma N$ channels, to reproduce the
observed pole position.  We will refer to this first scenario (Scenario I) as the ``bare Roper''
scenario.

In the second hypothesis, the Roper resonance is dynamically generated purely from the $\pi N$,
$\pi \Delta$, and $\sigma N$ channels.  We will refer to this second scenario (Scenario II) as the
``without bare baryon'' scenario.

In the third hypothesis, the Roper resonance is composed of the low-lying bare component associated
with the ground state nucleon.  Through coupling with the virtual meson-baryon contributions the
scattering data and pole position are reproduced.  We will refer to this third scenario (Scenario
III) as the ``bare Nucleon'' scenario.

We fit the experimental data up to centre of mass energies of 1800 MeV.  Since the $\sigma$ meson
has a large width and the $N(\pi\pi)_\texttt{S-wave}$ plays an important role for the
inelasticities of the $\pi N$ channel up to 1450 MeV, we fix the $\sigma$ mass to a small value of
350 MeV to describe the threshold behavior well.  
In this way we have included the threshold effects of the $N(\pi\pi)_\texttt{S-wave}$ channel in an
effective manner.  The contributions of both $N(\pi\pi)_\texttt{S-wave}$ and $N \sigma$ are
included in a single effective channel.  

Similar difficulties are encountered in lattice QCD.  Here one needs to include both five-quark
momentum-projected $N \sigma$ and seven-quark momentum-projected $N \pi \pi$ interpolating fields
to separate the $N(\pi\pi)_\texttt{S-wave}$ and $N \sigma$ contributions.  In the absence of the
seven-quark interpolating fields, the $N \pi \pi$ and $N \sigma$ contributions will be treated in a
similar effective manner, where the combined contributions are treated as a single state.

The fitted results for the phase shifts and inelasticities are plotted in Fig. \ref{figPSEta}, for
the three aforementioned scenarios.  The best-fit parameters and the pole positions for each
scenario are presented in Table \ref{tabPara}.  It is interesting to observe that a pole
corresponding to the Roper resonance is generated in all three scenarios, whether a bare state is
introduced or not.  While the imaginary part in Scenarios II and III deviates from the Review of
Particle Physics \cite{PDG2014}, we note the model is in agreement with the phase shift and
inelasticity data.

\begin{table}[tbp]
\caption{Best-fit parameters and resultant pole positions in the three scenarios: I, the system
  with the bare Roper; II, the system without a bare state; and III, the system with bare nucleon.
  Underlined parameters were fixed in the fitting of that scenario. The experimental pole position
  for the Roper resonance is $( 1365 \pm 15 ) - ( 95 \pm 15)\,i$ MeV \cite{PDG2014}.}
\label{tabPara}
\begin{ruledtabular}
\begin{tabular}{cccc}
\noalign{\smallskip}
Parameter                                    & I & II & III \\
\noalign{\smallskip}
\hline
\noalign{\smallskip}
$ g^S_{\pi N}                      $ & $0.161$             & $0.489$                   & $0.213$             \\
$ g^S_{\pi \Delta}                 $ & $-0.046$            & $-1.183$                  & $-1.633$            \\
$ g^S_{\pi N,\pi \Delta}           $ & $0.006$             & $-1.008$                  & $-0.640$            \\
$ g^S_{\pi N,\sigma N}             $ & $\underline{0}$     & $2.176$                   & $2.401$             \\
$ g^S_{\sigma N}                   $ & $\underline{0}$     & $9.898$                   & $9.343$             \\
$ g_{B_0\pi N}                     $ & $0.640$             & $\underline{0}$           & $-0.586$            \\
$ g_{B_0\pi\Delta}                 $ & $1.044$             & $\underline{0}$           & $1.012$             \\
$ g_{B_0\sigma N}                  $ & $2.172$             & $\underline{0}$           & $2.739$             \\
$ m_B^0/{\rm GeV}                  $ & $2.033$             & $\underline{\infty}$      & $1.170$             \\
$ \Lambda_{\pi N}/{\rm GeV}        $ & $\underline{0.700}$ & $0.562$                   & $\underline{0.562}$ \\
$ \Lambda_{\pi \Delta}/{\rm GeV}   $ & $\underline{0.700}$ & $0.654$                   & $\underline{0.654}$ \\
$ \Lambda_{\sigma N}/{\rm GeV}     $ & $\underline{0.700}$ & $1.353$                   & $\underline{1.353}$ \\
\noalign{\medskip}
Pole  (MeV)                          & $1380 - 87\,i$      & $1361 - 39\,i$            & $1357 - 36\,i$      \\
\end{tabular}
\end{ruledtabular}
\end{table}

\begin{figure}[t]
\begin{center}
\includegraphics[width=1.0\columnwidth]{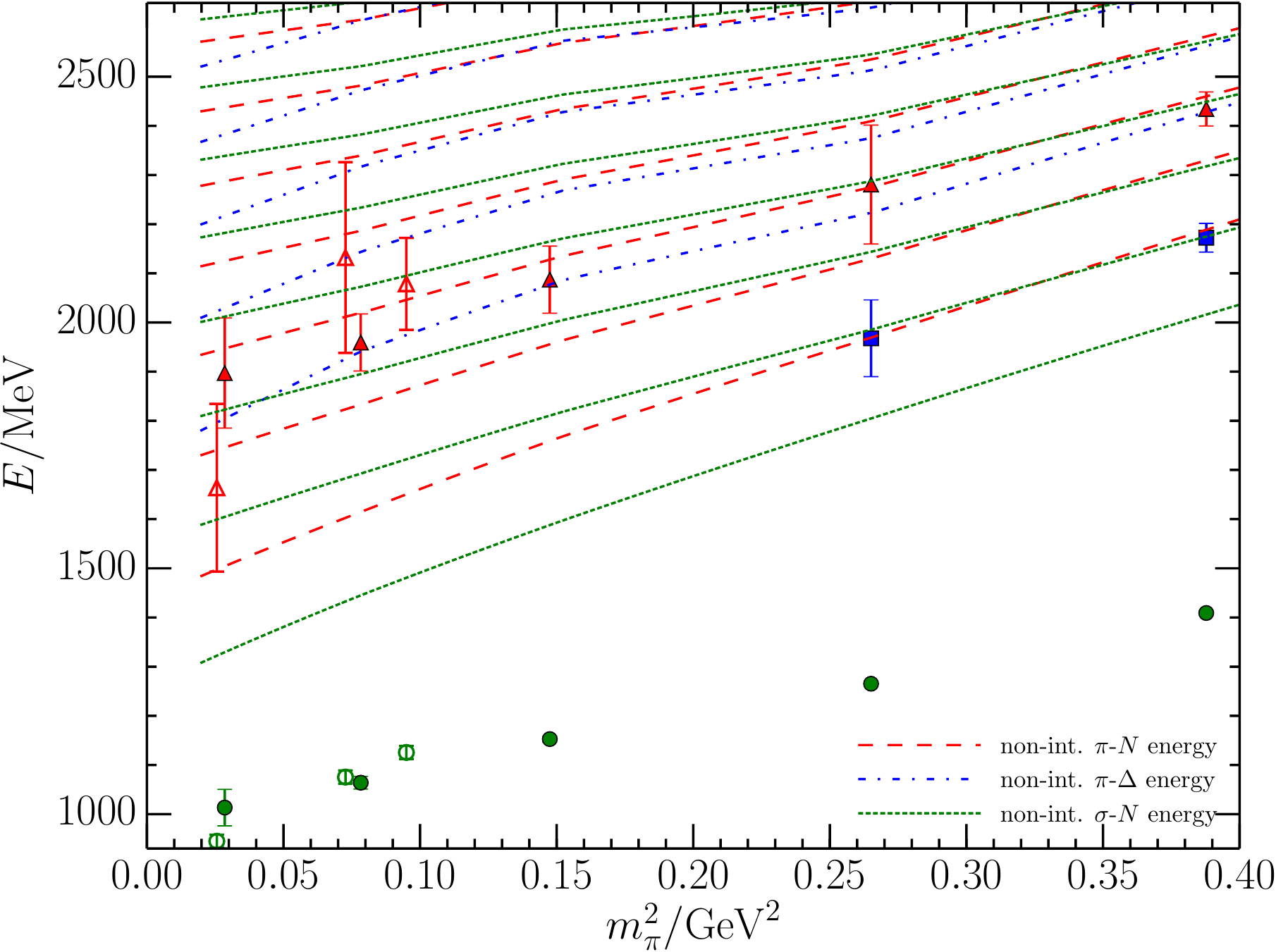}
\caption{{\bf Colour online:} The lowest lying $I(J^P)=\frac12(\frac12^+)$ baryon states observed
  in lattice QCD simulations with length $L \simeq 2.90$ fm, as a function of the input quark mass
  ($\propto m_\pi^2$).  The data with filled symbols are from the CSSM group
  \cite{Mahbub:2013ala,Mahbub:2010rm,Kiratidis2015} updated to high statistics herein.  Those with
  hollow symbols are from the Cyprus group \cite{Alexandrou2015}.  The non-interacting energies of
  the low-lying two-particle meson-baryon channels for this lattice are also illustrated. }
\label{figLDNonInt3fm}
\end{center} 
\end{figure}

With the parameters of the interactions constrained by the experimental phase shifts and
inelasticities, one can proceed to compare the predictions of the matrix Hamiltonian model in a
finite volume with results from lattice QCD.  Fig.~\ref{figLDNonInt3fm} summarises world lattice
QCD results \cite{Leinweber:2015kyz} for the positive parity nucleon spectrum
\cite{Mahbub:2013ala,Mahbub:2010rm,Kiratidis2015,Alexandrou2015} for volumes with $L \simeq 2.9$
fm.  Here the statistics of the CSSM results \cite{Mahbub:2013ala} have been increased through the
consideration of approximately 29,472 propagators on the PACS-CS configurations \cite{Aoki:2008sm}.
The $2S$ orbital structure of the CSSM's first excited states of the nucleon reported in
Fig.~\ref{figLDNonInt3fm} was established in Refs.~\cite{Roberts:2013ipa} and
\cite{Roberts:2013oea}.

It may be interesting to note that the CSSM collaboration performed a rather exhaustive search for
a low-lying Roper-like state in Ref.~\cite{Mahbub:2010rm}.  There a broad range of smeared-source
interpolators were considered in the hope that one could form a correlation matrix that would
reveal a low-lying Roper-like state.  Instead, one found that the first excitation energy was
insensitive to the basis of interpolating fields explored and no state approaching 1440 MeV could
be found.  

It is important to note that these lattice QCD results have been obtained through the use of local
three-quark interpolating fields.  This approach will make it difficult to access multi-particle
scattering states.  If there is little attraction to localize the multi-particle state in a
finite volume, $V$, then the overlap of the interpolator is volume-suppressed by a factor of $1/V$;
{\it i.e} the probability of finding the second hadron at the position of the first is $1/V$.
As a consequence, it may be that only the states composed of a significant bare-state component in
the Hamiltonian model will be excited by the three-quark interpolating fields.  We will examine
this possibility in detail in the following.

The non-interacting energies of the two-particle meson-baryon channels considered herein for
$L = 2.90$ fm are also illustrated in Fig.~\ref{figLDNonInt3fm}.  The predictions of the three
scenarios for the finite-volume spectra of lattice QCD are presented in Secs.\ \ref{secSBR},
\ref{secSNB} and \ref{secSBN}.

\subsection{System with the Bare Roper Basis State} \label{secSBR}

In the traditional quark model, the nucleon is thought to be made up of three constituent quarks
with small five-quark components \cite{Julia-Diaz2006}.  In the view of effective field theory, the
nucleon is a mixed state of a bare nucleon component, dressed by attractive $\pi N$, $\pi \Delta$,
etc.\ components, with the bare component dominating.  As a simple analogy, the Roper is popularly
treated as a state dominated by a bare Roper component with a mass above the Roper resonance
position, dressed by attractive $\pi N$, $\pi \Delta$, etc.\ components.  The bare Roper component
is thought to coincide with a three-quark core while other components like $\pi N$ contain states
with at least five quarks.

Our first scenario follows this picture.  In this scenario the Roper is composed of a large-mass
bare Roper state dressed by $\pi N$, $\pi \Delta$, and $\sigma N$ channels.  Fits to the
phase shifts and inelasticities in this model are plotted as the red dot-dashed lines in
Fig.~\ref{figPSEta}, and the associated parameters are listed in the column labeled I in Table
\ref{tabPara}.  In this fit, two of the separable potentials are removed via $g^S_{\sigma N,\sigma
  N}=0$ and $g^S_{\pi N,\sigma N}=0$, because their effect on the fits is accommodated by other
interactions.

\begin{figure}[t]
\begin{center}
\includegraphics[width=1.0\columnwidth]{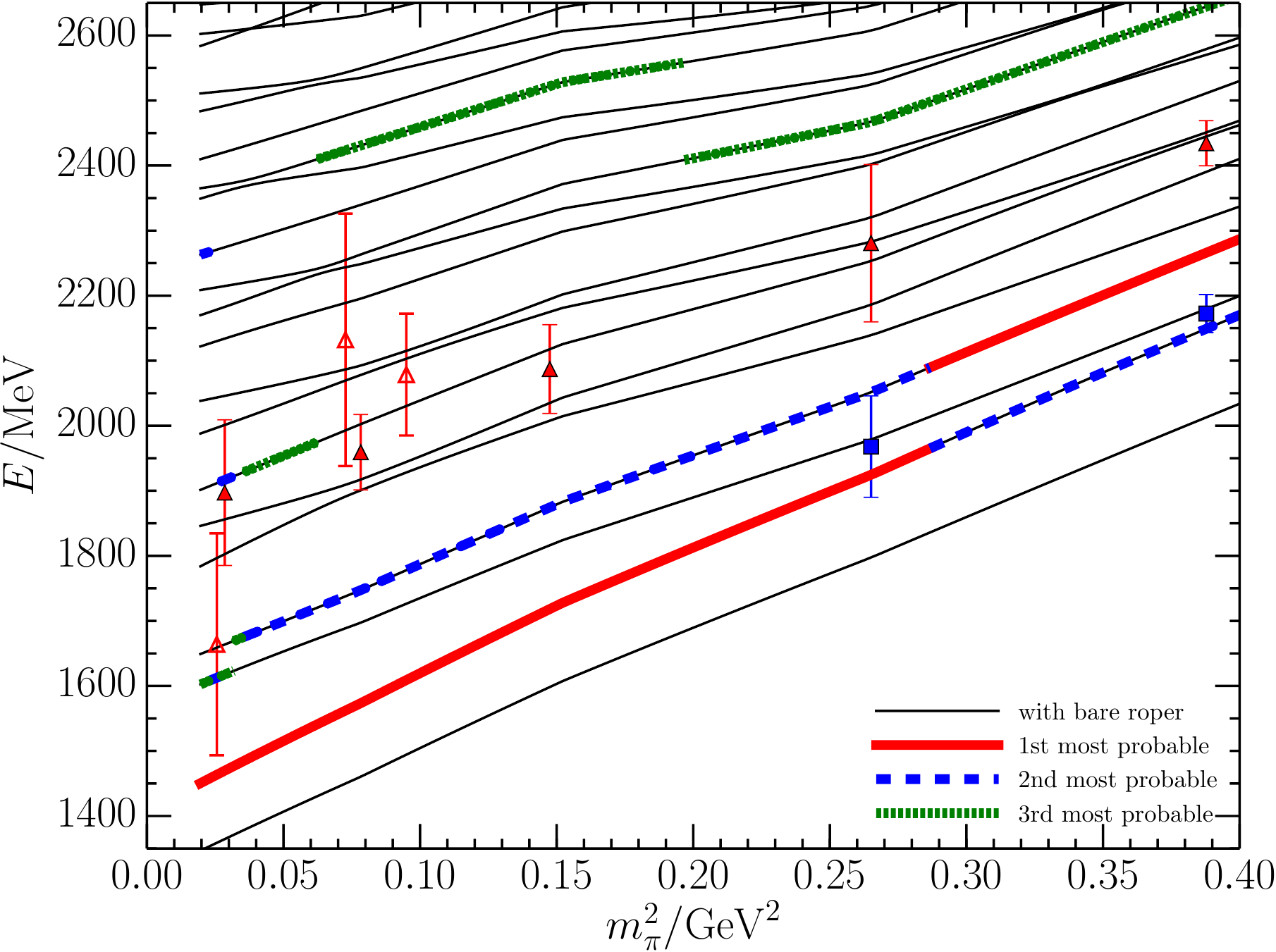}
\caption{{\bf Colour online}: The pion mass dependence of the $L = 2.90$ fm finite-volume
  energy eigenstates for the Hamiltonian model scenario with a bare Roper basis state.  The different
  line types and colours used in illustrating the energy levels indicate the strength of the bare
  basis state in the Hamiltonian-model eigenvector describing the composition of the state.  The
  thick-solid (red), dashed (blue) and dotted (green) lines correspond to the states having the
  first, second, and third largest bare-state contributions.  Since three-quark operators are used
  to excite the states observed in lattice QCD, we label these states as the first, second, and
  third most probable states to be seen in the lattice QCD simulations.}
\label{figSpecDpl3fm}
\end{center} 
\end{figure}

These fit parameters enable the determination of the eigen-energy spectra in a finite volume at the
physical pion mass.  To obtain the spectrum at higher quark masses, we proceed to determine the
quark mass dependence of the bare mass, $m_R^0(m_\pi^2)$, governed by the slope parameter
$\alpha_R^0$.  To do this, we consider the lowest-lying excitation energies observed by the CSSM at
the largest two quark masses considered, illustrated by the filled blue-square markers in
Fig.~\ref{figLDNonInt3fm}.  We assume that these lattice results, obtained with three-quark
operators, correspond to Hamiltonian-model states in which the bare state plays an important role.
The parameter $\alpha_R^0$ is then constrained via a standard $\chi^2$ measure between the first
Hamiltonian model excitation with a significant bare state eigenvector component the aforementioned
lattice QCD results.  The remainder of the spectra are then a prediction.  Our best fit gives
$\alpha_R^0=2.14~{\rm GeV^{-1}}$.

The energy levels of the Hamiltonian model for the lattice with $L = 2.90$ fm are illustrated
in Fig.~\ref{figSpecDpl3fm}.  In this case, the lowest lying excitation in the Hamiltonian model is
almost pure $\sigma N$.  The second excitation has a bare state component exceeding 10\% and
therefore this state is constrained to the lowest-lying excitation energies at the largest two
quark masses considered, illustrated by the filled blue-square markers in
Fig.~\ref{figLDNonInt3fm}, in determining $\alpha_R^0$.

\begin{figure}[t]
\begin{center}
\includegraphics[width=1.0\columnwidth]{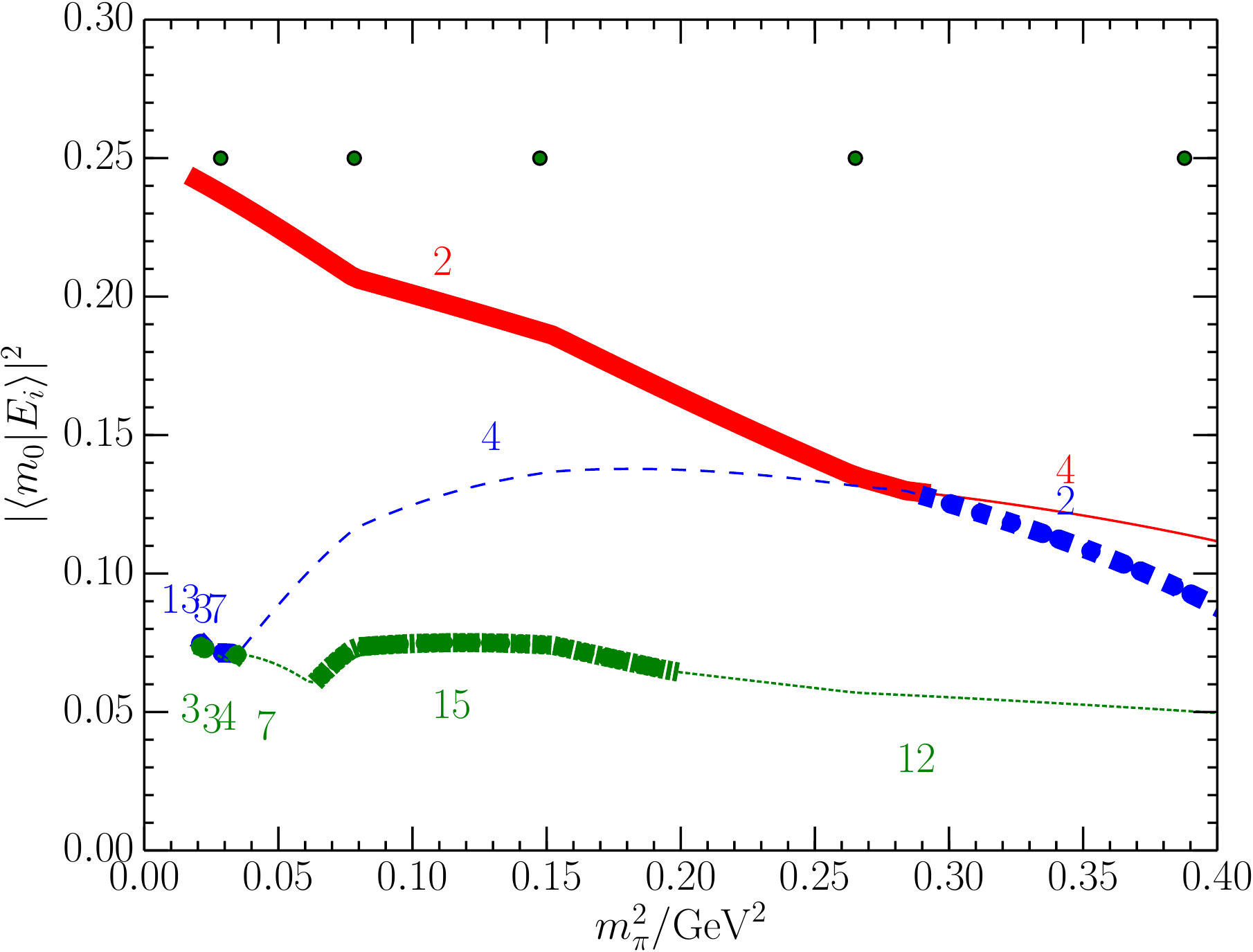}
\caption{{\bf Colour online:} The fraction of the bare-Roper basis state, $| m_0 \rangle$, in the
  Hamiltonian energy eigenstates $| E_i \rangle$ for the three states having the largest bare-state
  contribution.  States are labeled by the energy-eigenstate integers $i$ and these state labels
  are indicated next to the curves.  For example, at light quark masses, the second energy
  eigenstate has the largest bare-Roper component and therefore the second excitation energy in
  Fig.~\ref{figSpecDpl3fm} is highlighted with a thick red line.  The dark-green dots plotted at
  $y = 0.25$ indicate the positions of the five quark masses considered in the CSSM results.  While
  the line type and colour scheme match that of Fig.~\ref{figSpecDpl3fm}, the thick and thin
  lines alternate to indicate a change in the energy eigenstate associated with that value.}
\label{figBareRDpl3fm}
\end{center} 
\end{figure}

The (coloured) thick-solid, dashed and dotted line-type decorations of the Hamiltonian model
eigenstates in Fig.~\ref{figSpecDpl3fm} reflect the magnitude of the bare-state contribution to
the eigenstates.
Denoting $|E_i\rangle$ as the $i$-th energy eigenstate from the matrix Hamiltonian model the
structure of $|E_i\rangle$ is obtained through the overlap of the eigenvector with each of the
basis states.  For example, the proportion of the bare state $|m_0\rangle$ in $|E_i\rangle$ is
$|\langle m_0|E_i\rangle|^2$.  For the meson-baryon basis states, we sum over all the back-to-back
momenta considered when reporting their contributions to the energy eigenstates.

Since three-quark operators are used to excite the states observed in lattice QCD, Hamiltonian
model states with a large proportion of the bare-Roper basis state are more likely to be observed
in the lattice QCD calculations.  To identify these states, we seek the first three eigenstates,
$|E_i\rangle$, among the first twenty lowest-lying excitations, which contain the largest
bare-Roper basis state contributions.  This is done at each pion mass considered in generating the
curves.  We label these states in Fig.~\ref{figSpecDpl3fm} as the first, second, and third most
probable states to be seen in the lattice QCD simulations.

Fig.~\ref{figBareRDpl3fm} reports the bare-Roper fraction, $|\langle m_0|E_i\rangle|^2$, for these
three states.  The integer next to each section of the curves indicates the $i$-th energy
eigenstate associated with the fraction plotted.
We see that the bare-Roper basis state strength is spread across many energy eigenstates.  None of
the first twenty eigenstates contain more than 30\% of the state in the bare-Roper basis
state. This situation contrasts that for the nucleon ground state where more than 80\% of the
energy eigenstate is composed with the bare-nucleon basis state.

\begin{figure}[t]
\begin{center}
\subfigure[\ 1st eigenstate]{\scalebox{\sca}{\includegraphics{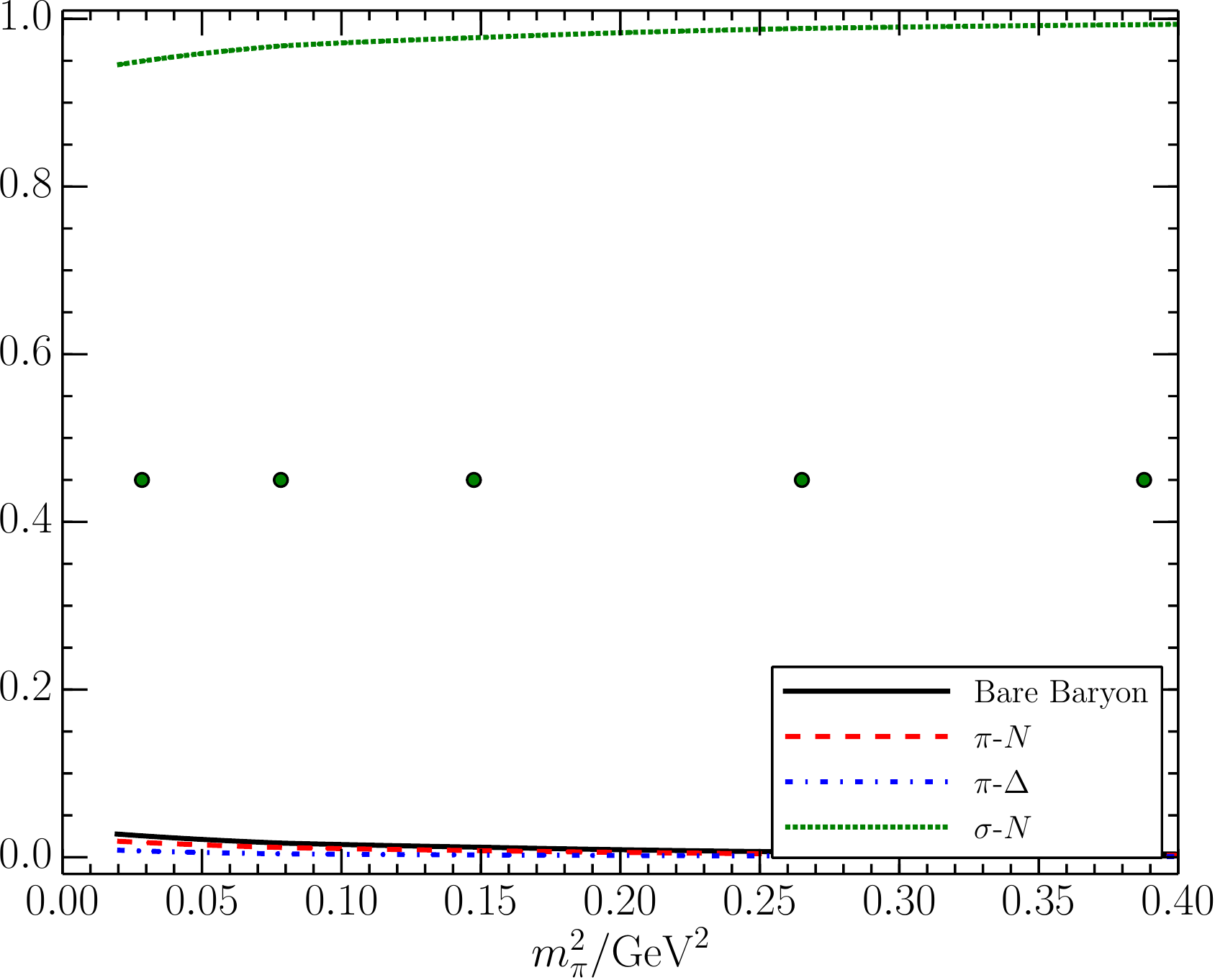}}\label{figComDpl3fm1}}
\subfigure[\ 2nd eigenstate]{\scalebox{\sca}{\includegraphics{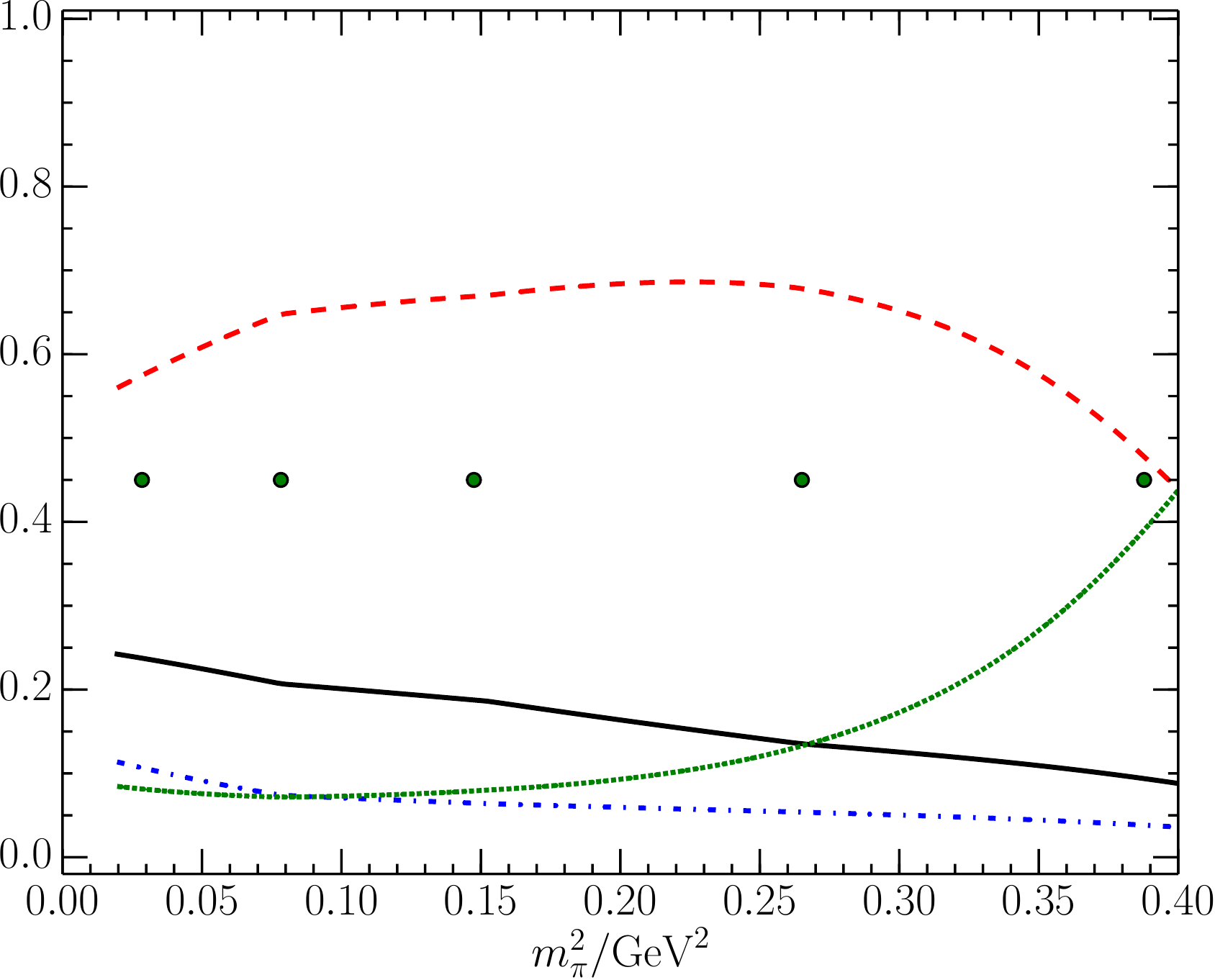}}\label{figComDpl3fm2}}
\subfigure[\ 3rd eigenstate]{\scalebox{\sca}{\includegraphics{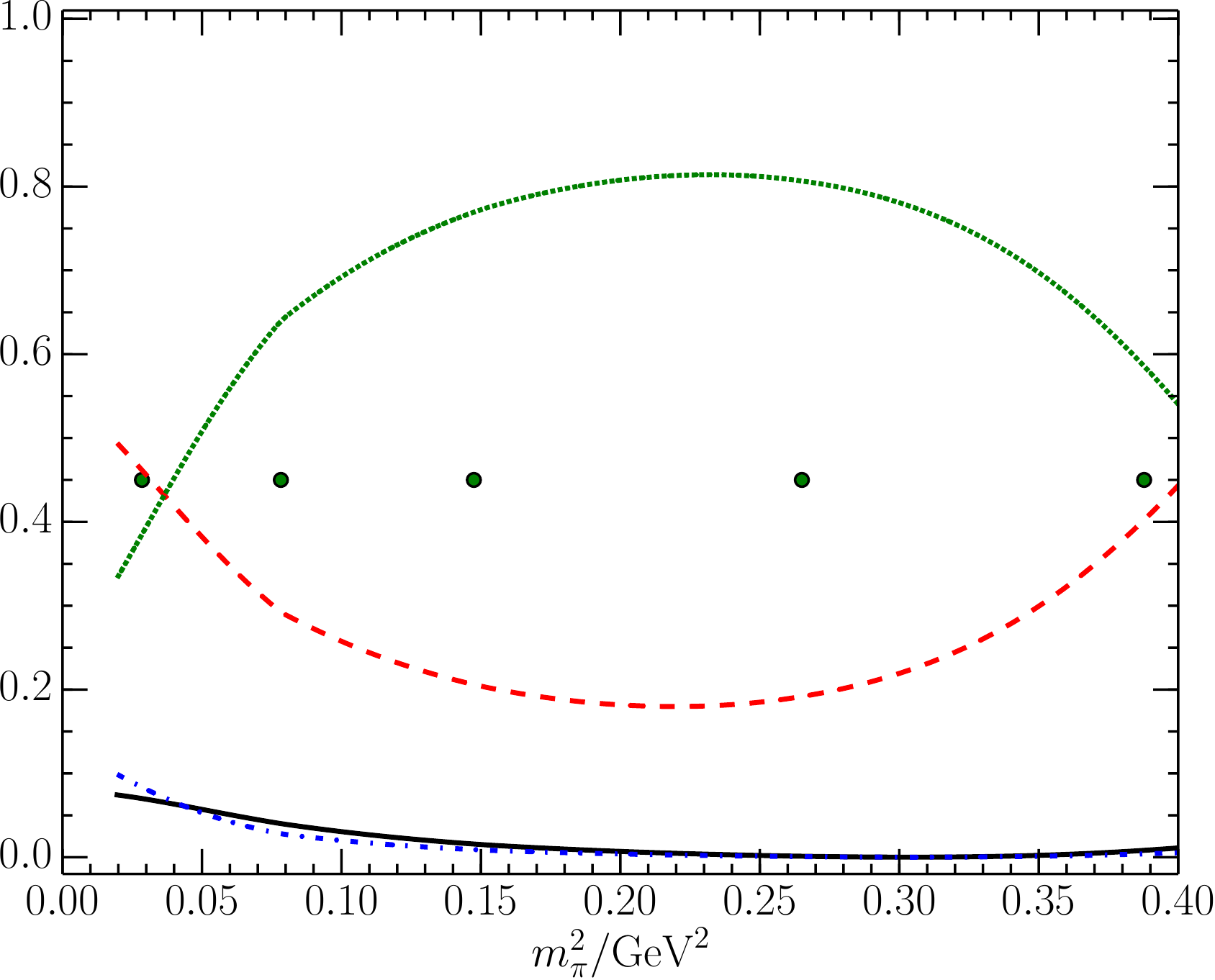}}}
\subfigure[\ 4th eigenstate]{\scalebox{\sca}{\includegraphics{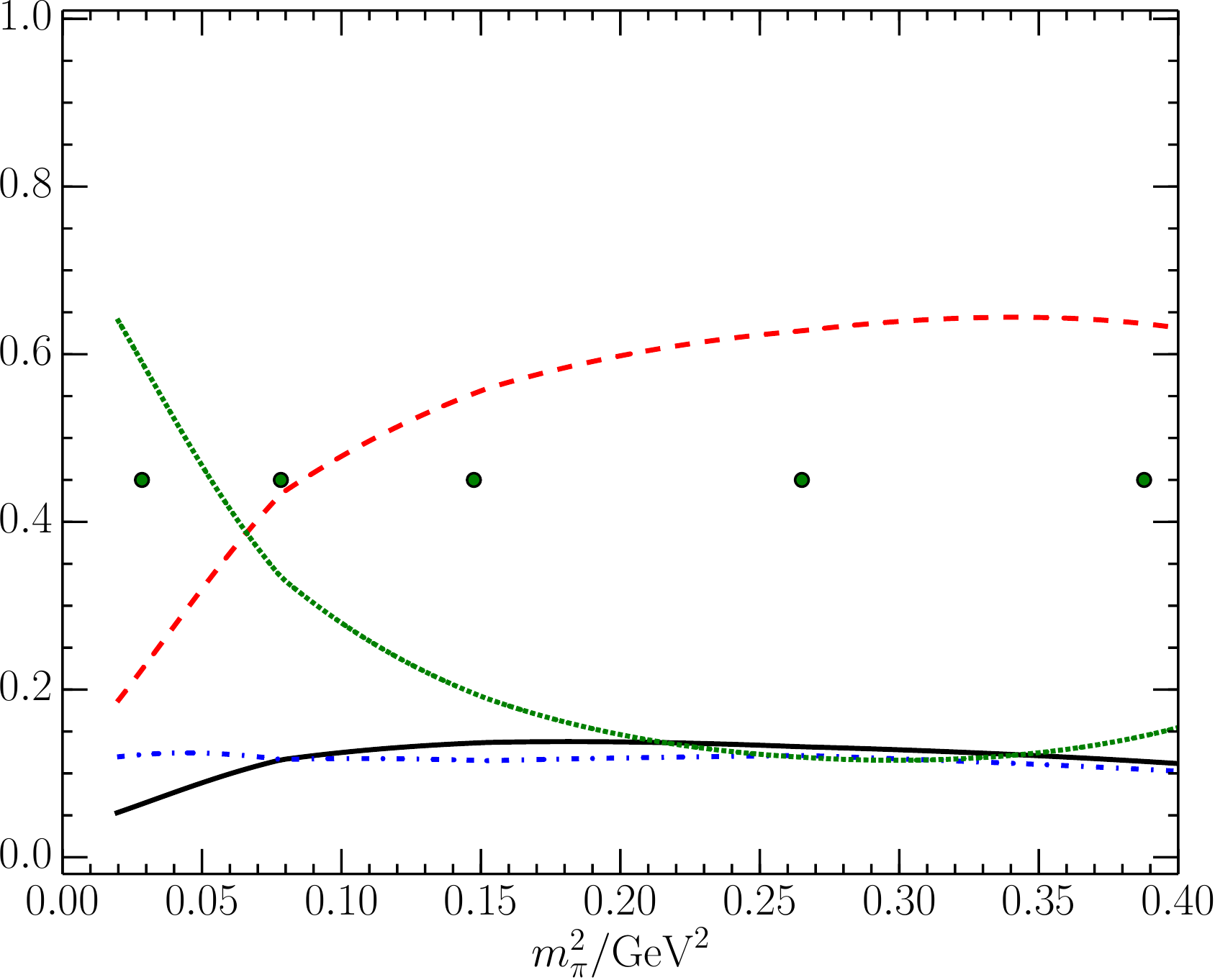}}}
\caption{{\bf Colour online:} The pion-mass evolution of the Hamiltonian eigenvector components for
  the scenario with the bare Roper on the lattice volume with $L = 2.90$ fm.  The fractions of
  the bare-state, $|\langle m_0|E_i\rangle|^2$, and meson-baryon channels,
  $\sum_{\vec k} |\langle \alpha(\vec k) | E_i\rangle|^2$, composing the energy
  eigenvectors are illustrated for the first four states observed in the model.  Here all momenta
  for a particular meson-baryon channel have been summed to report the relative importance of the
  $\alpha = m_0$, $\pi N$, $\pi \Delta$ and $\sigma N$ channels. The (green) dots plotted horizontally at $y
  = 0.45$ indicate the positions of the five pion masses considered by the CSSM.
\vspace{-12pt}}
\label{figCompnDpl3fm}
\end{center} 
\end{figure}

To further illustrate the composition of the energy eigenstates created in this scenario,
Fig.~\ref{figCompnDpl3fm} reports the fractions of the bare-state and meson-baryon channels
composing the energy eigenvectors as a function of the squared pion mass for the first four
low-lying states in the finite volume with $L = 2.90$ fm.
The first panel, Fig.~\ref{figComDpl3fm1}, shows the first state is nearly a pure $|\sigma N\rangle$
scattering state associated with $N(\pi\pi)_\texttt{S-wave}$ contributions.

Fig.~\ref{figComDpl3fm2} reveals the second lowest state is dominated by the $|\pi N\rangle$ basis
state, as argued in the original publications reporting this state
\cite{Mahbub:2013ala,Mahbub:2010rm}.  Away from the avoided level crossing at the largest quark
mass considered, the state is typically 60\% $|\pi N\rangle$ 20\% bare Roper, 10\% $|\pi
\Delta\rangle$ and 10\% $|\sigma N\rangle$.  The presence of a significant bare-state contribution
explains the ability of the three-quark interpolating fields used in the lattice QCD calculations
to excite this state.  Similarly the absence of a significant bare-state contribution to the first
excitation explains the omission of this state in present-day lattice QCD simulations.

Of particular note is the prediction of a very significant bare-Roper contribution to the second
energy eigenstate at light quark masses.  The bare-state contribution exceeds 15\% for the three
lightest quark masses considered by the CSSM, {\it i.e.}  $0 \le m_\pi^2 \le 0.15\ {\rm GeV^2}$.
Therefore this scenario predicts that these low-lying states should be readily observed in the
lattice QCD calculations.  However, neither the CSSM nor the Cyprus groups have observed these
states.  Therefore, this scenario in inconsistent with lattice QCD.  Thus the popular notion of the
Roper resonance being described by a large bare-Roper mass dressed by attractive meson-baryon
scattering channels is not supported by lattice QCD.

\subsection{System Without a Bare Baryon Basis State}\label{secSNB}

In light of the discrepancy between the first scenario and the results of lattice QCD, we proceed
to explore the possibility that Roper resonance is a pure molecular state.  In this scenario, the
Roper is assumed to be void of any triquark core and therefore we do not introduce a bare-baryon
basis state.  If this model can describe the experimental data, then it can also explain the void
of low-lying states in lattice QCD, as the overlap of three-quark operators with multi-particle
states is volume suppressed.

\begin{figure}[t]
\begin{center}
\includegraphics[width=1.0\columnwidth]{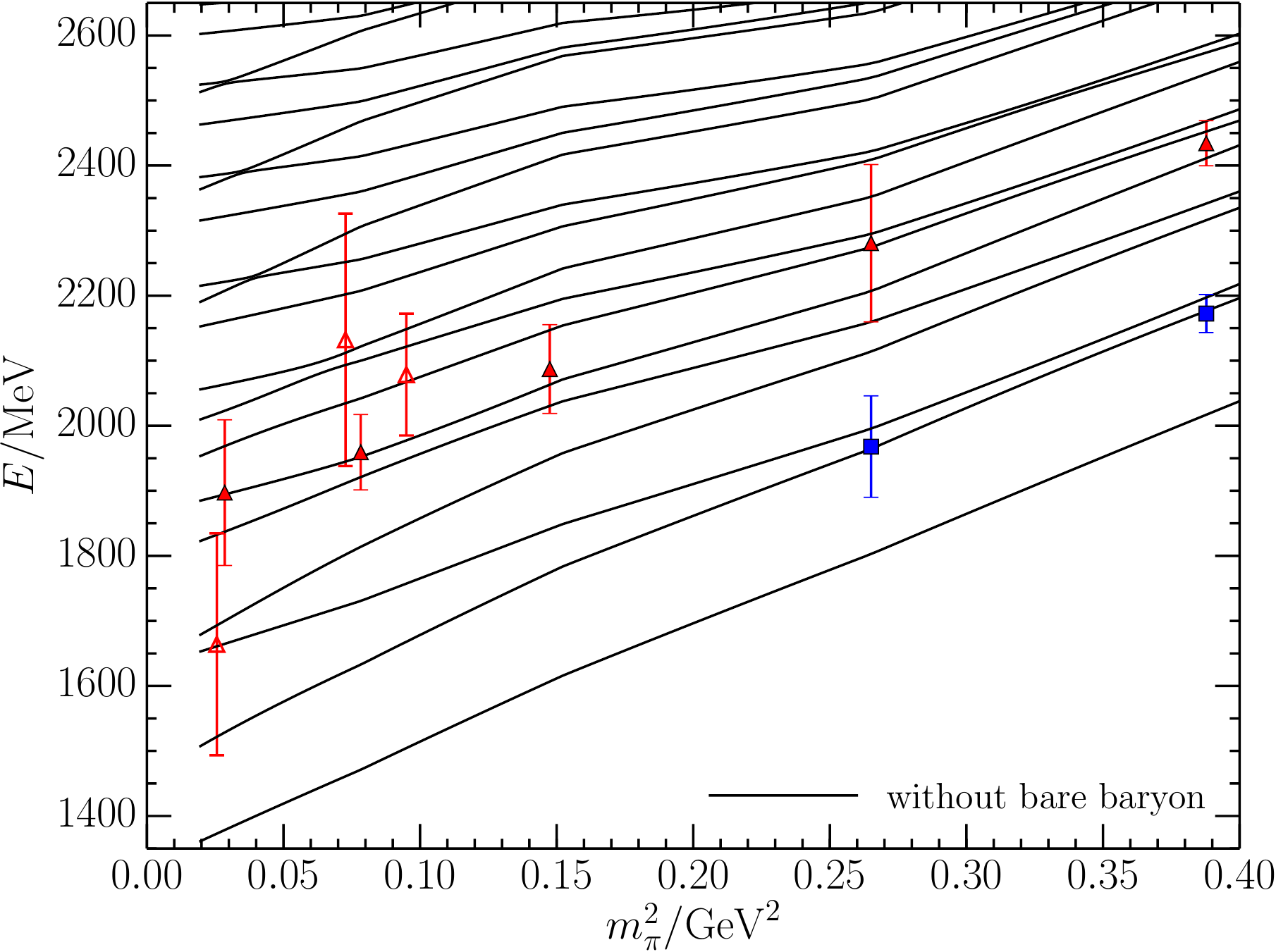}
\caption{{\bf Colour online:} The pion mass dependence of the $L = 2.90$ fm finite-volume
  energy eigenstates for the Hamiltonian model scenario without a bare baryon basis state.}
\label{figSpecExp3fm}
\end{center} 
\end{figure}

The fitted phase shifts and inelasticities are plotted as dotted lines in Fig.~\ref{figPSEta}
indicating the scattering data can be fit in the absence of a bare baryon contribution.  The
corresponding fit parameters are reported in the middle column of Table \ref{tabPara}.
Fig.~\ref{figSpecExp3fm} displays the energy levels in the finite-volume lattice.  The high density
of eigenstate levels from the Hamiltonian model provides easy overlap with the lattice QCD results.

\begin{figure}[tb]
\begin{center}
\subfigure[\ 1st eigenstate]{\scalebox{\sca}{\includegraphics{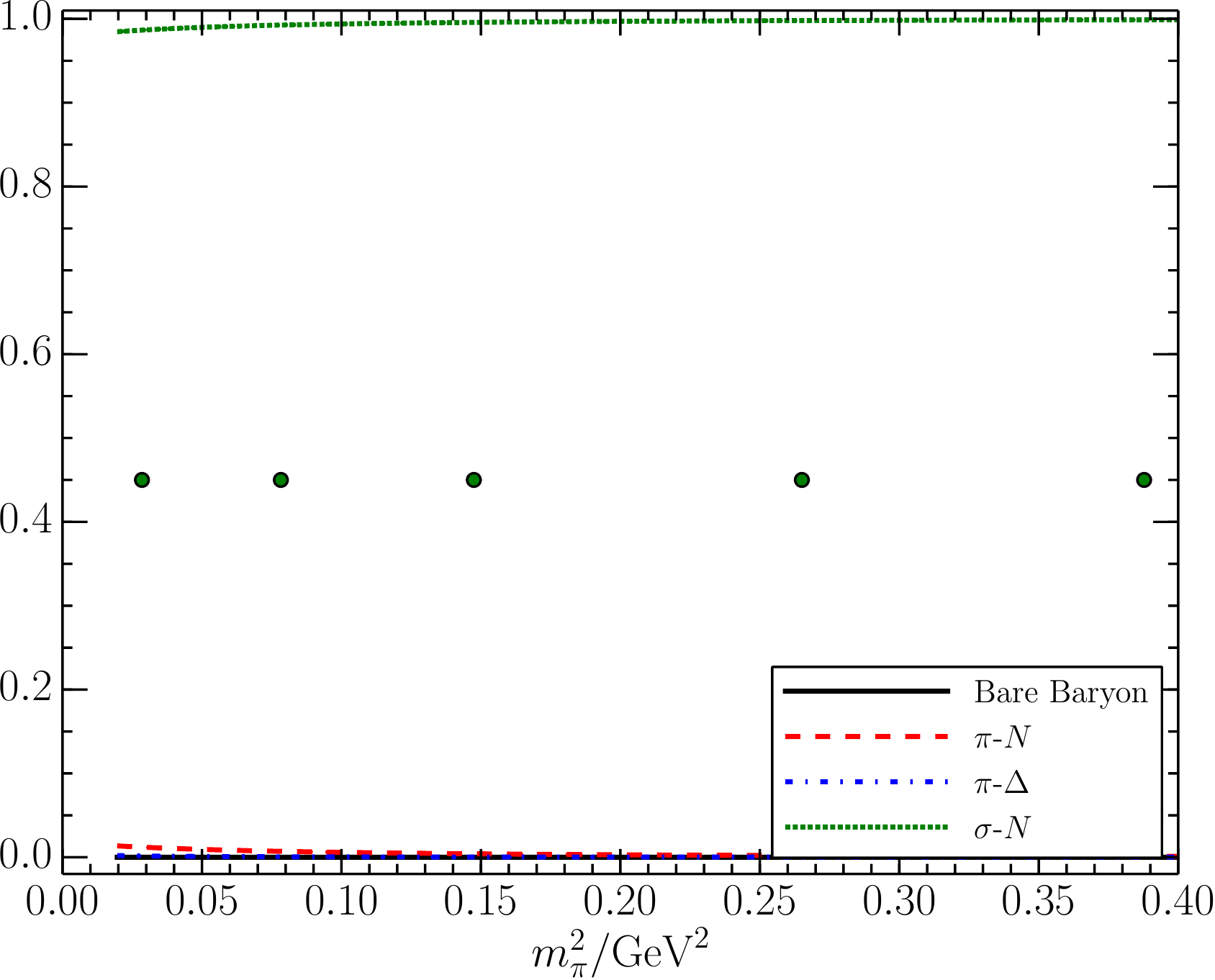}}\label{figComExp3fm1}}
\subfigure[\ 2nd eigenstate]{\scalebox{\sca}{\includegraphics{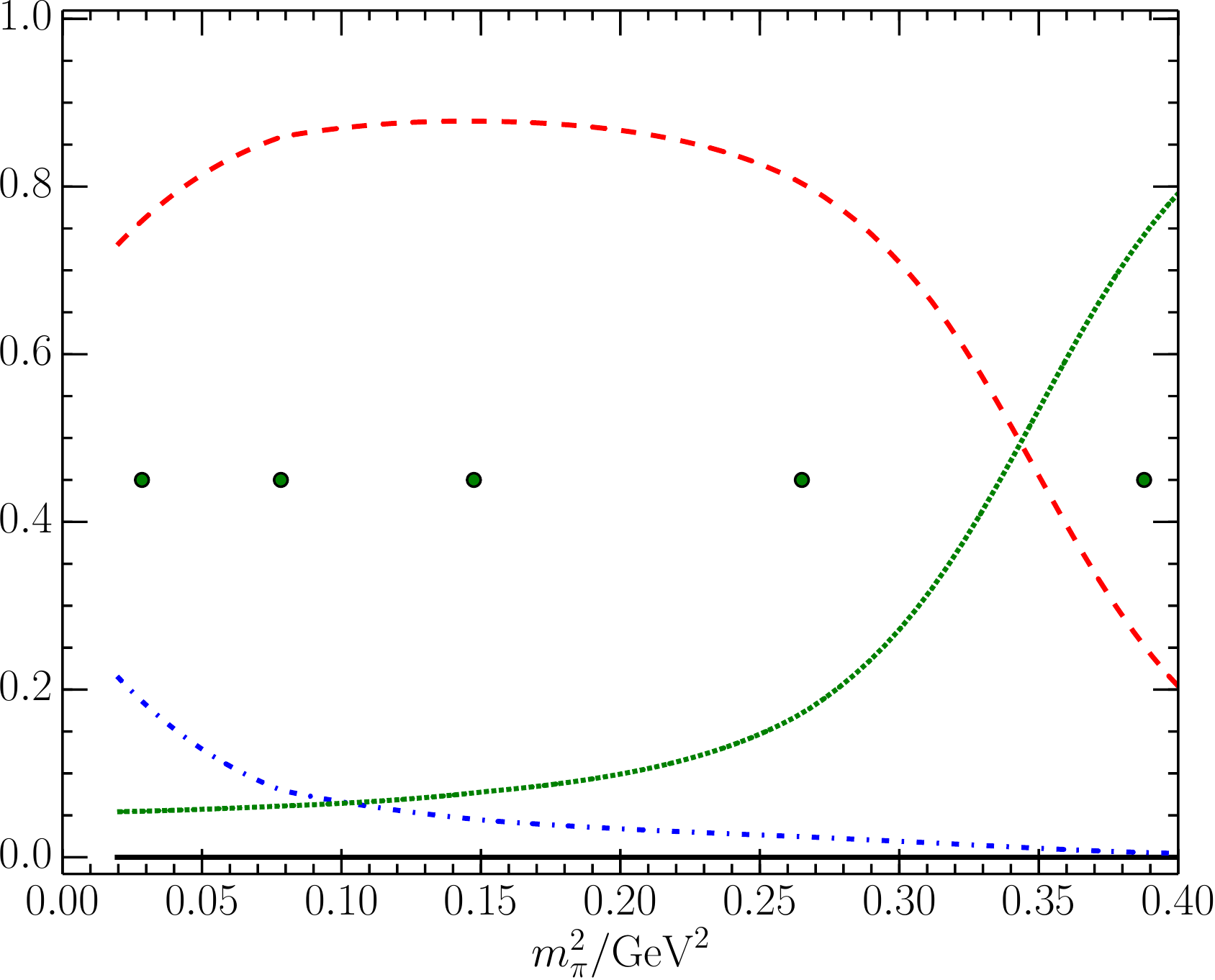}}\label{figComExp3fm2}}
\subfigure[\ 3rd eigenstate]{\scalebox{\sca}{\includegraphics{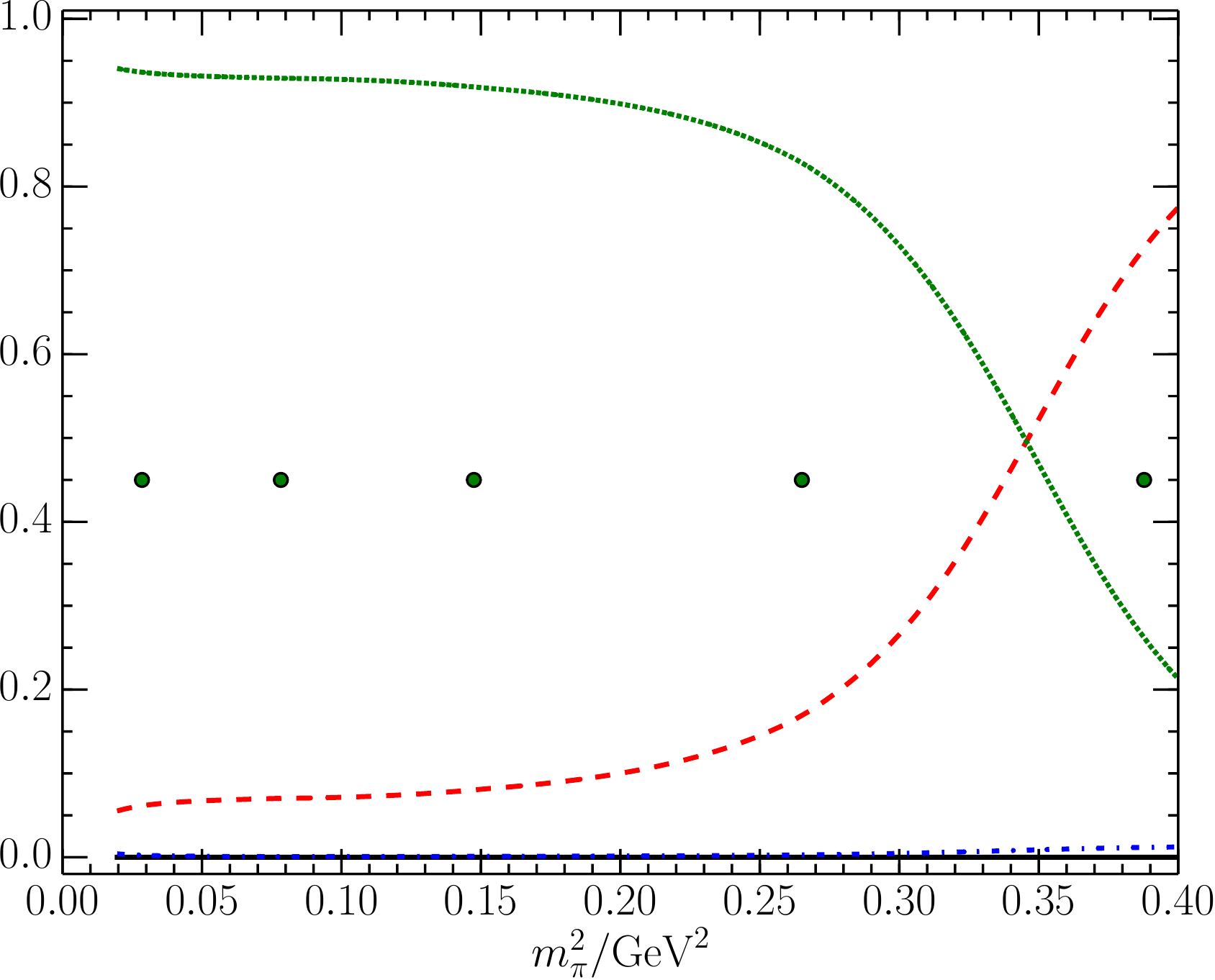}}\label{figComExp3fm3}}
\subfigure[\ 4th eigenstate]{\scalebox{\sca}{\includegraphics{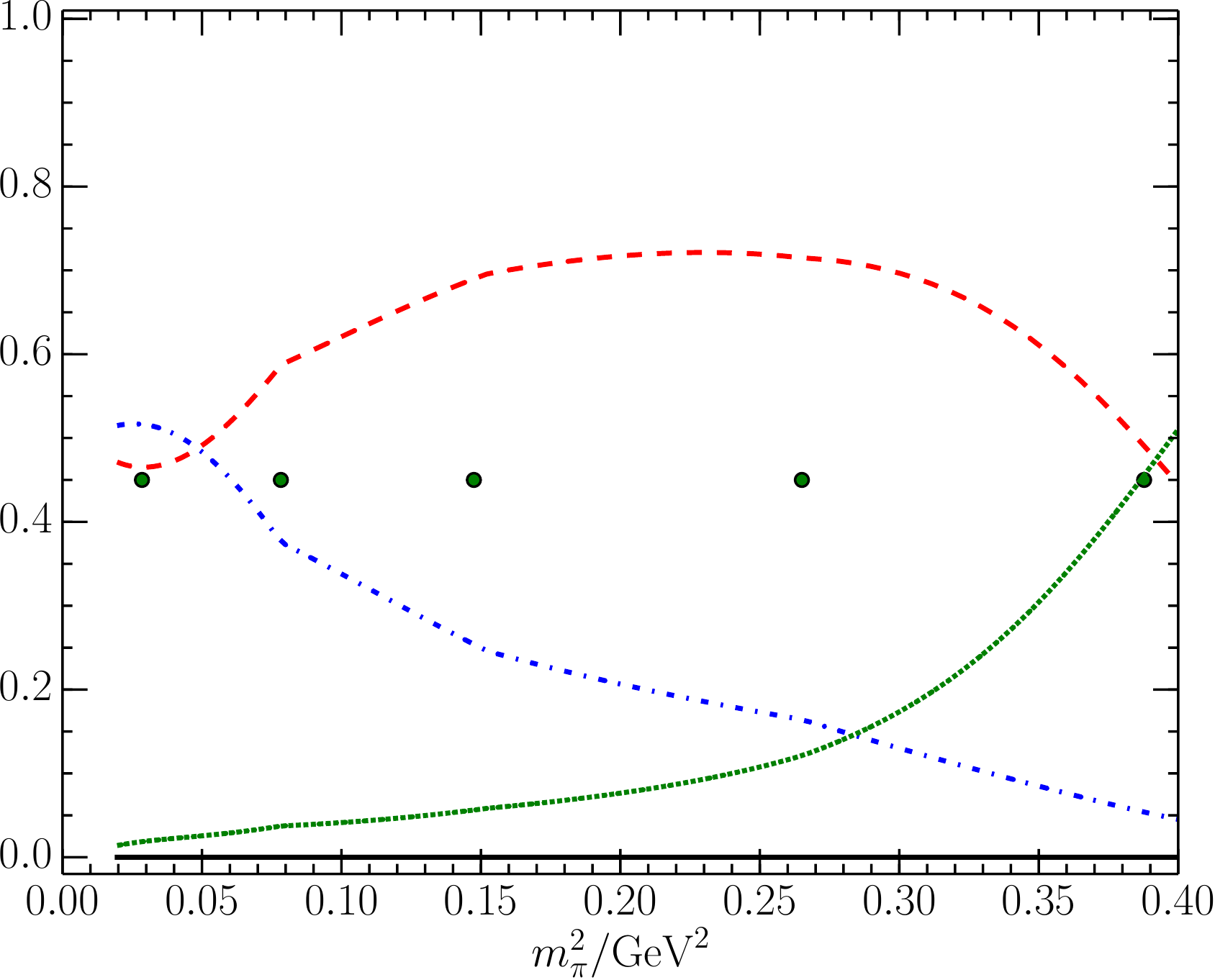}}\label{figComExp3fm4}}
\subfigure[\ 5th eigenstate]{\scalebox{\sca}{\includegraphics{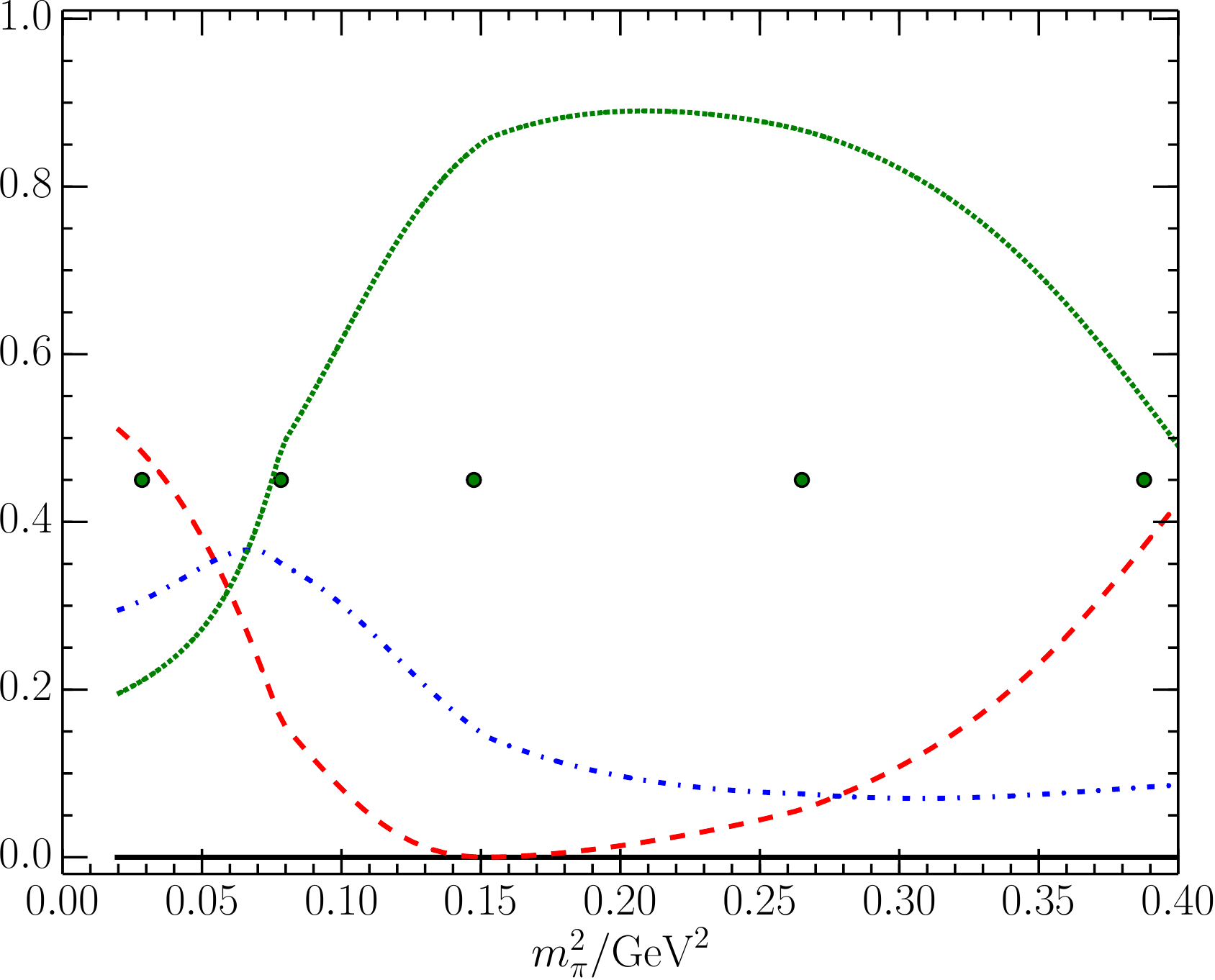}}\label{figComExp3fm5}}
\subfigure[\ 6th eigenstate]{\scalebox{\sca}{\includegraphics{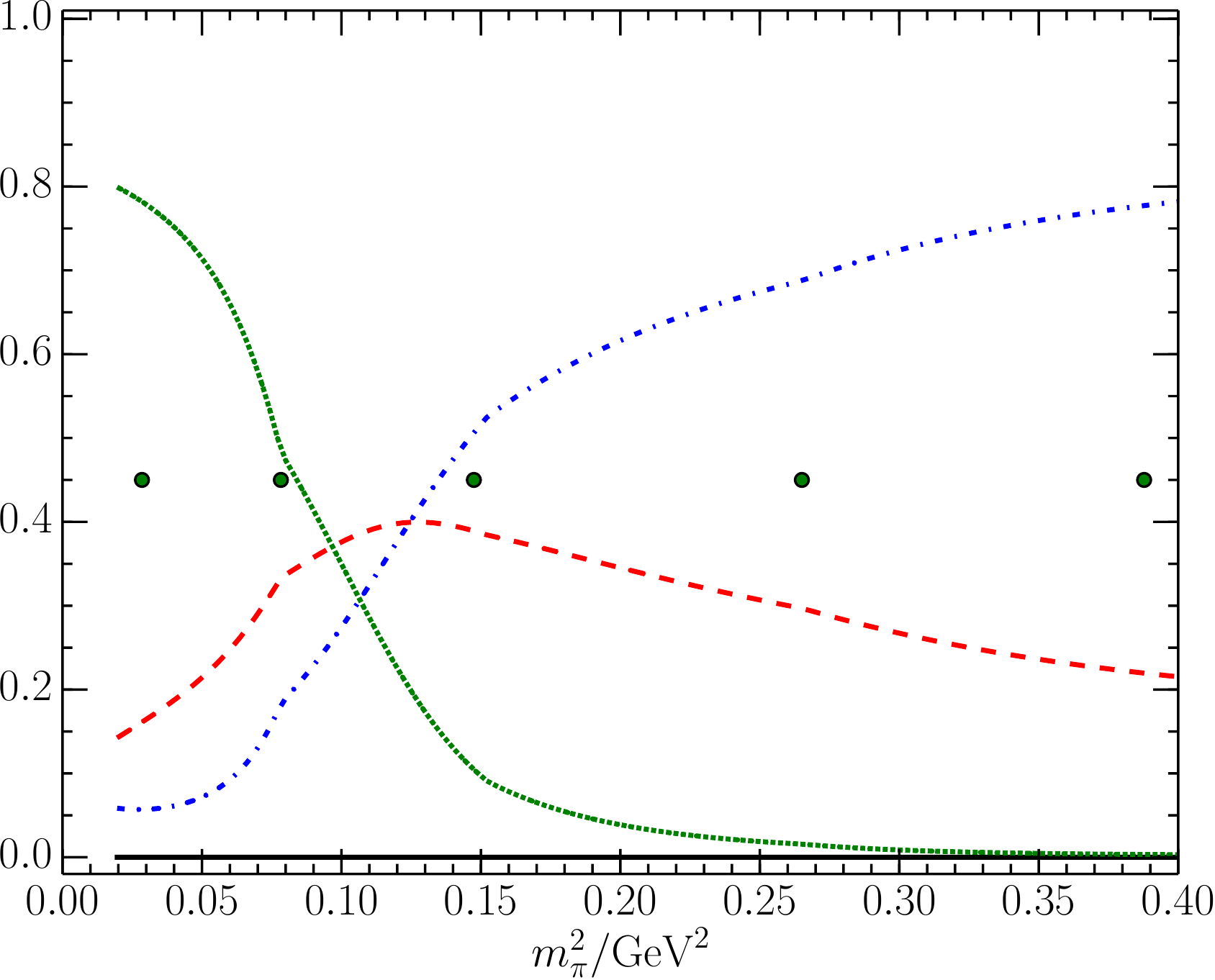}}\label{figComExp3fm6}}
\subfigure[\ 7th eigenstate]{\scalebox{\sca}{\includegraphics{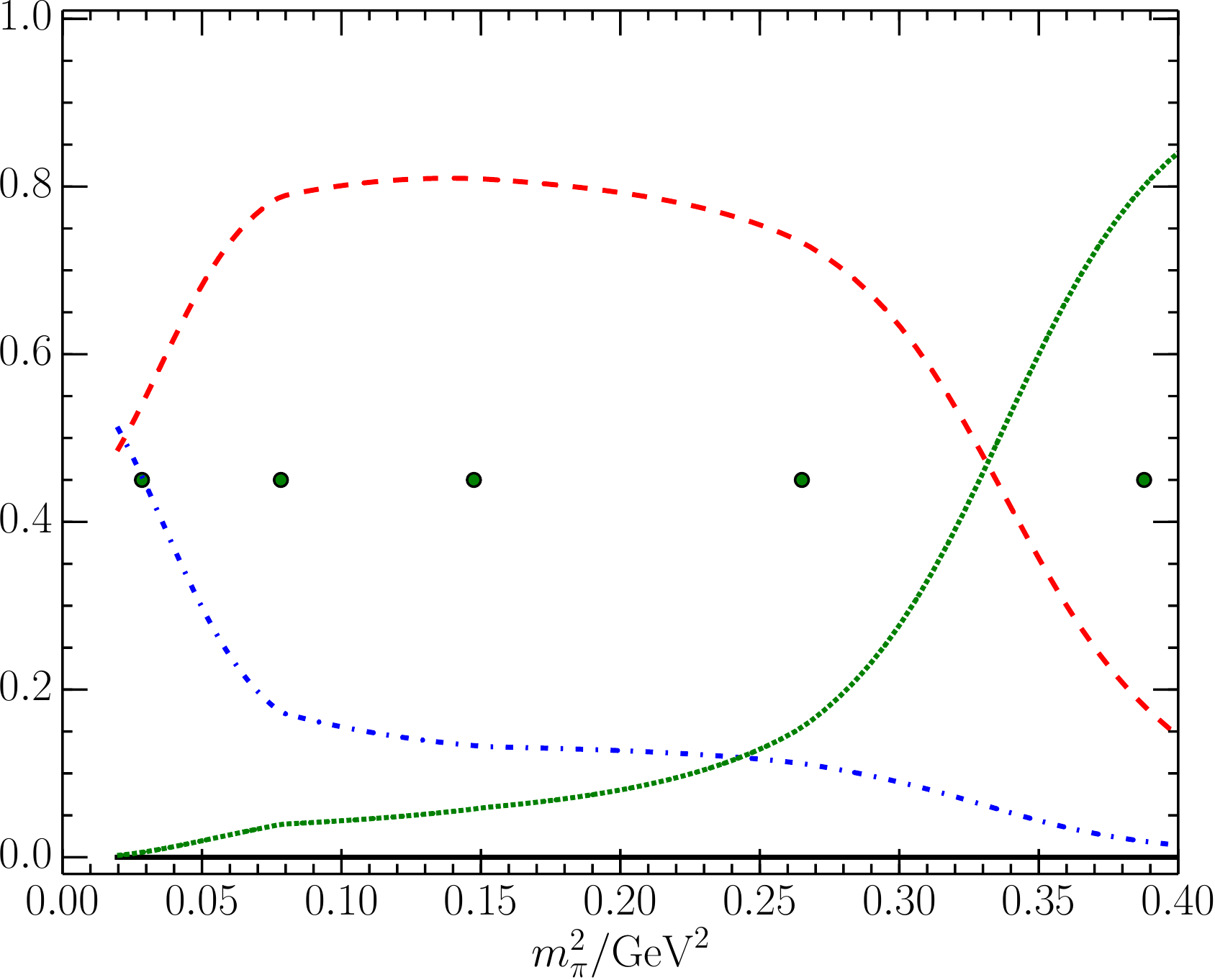}}\label{figComExp3fm7}}
\subfigure[\ 8th eigenstate]{\scalebox{\sca}{\includegraphics{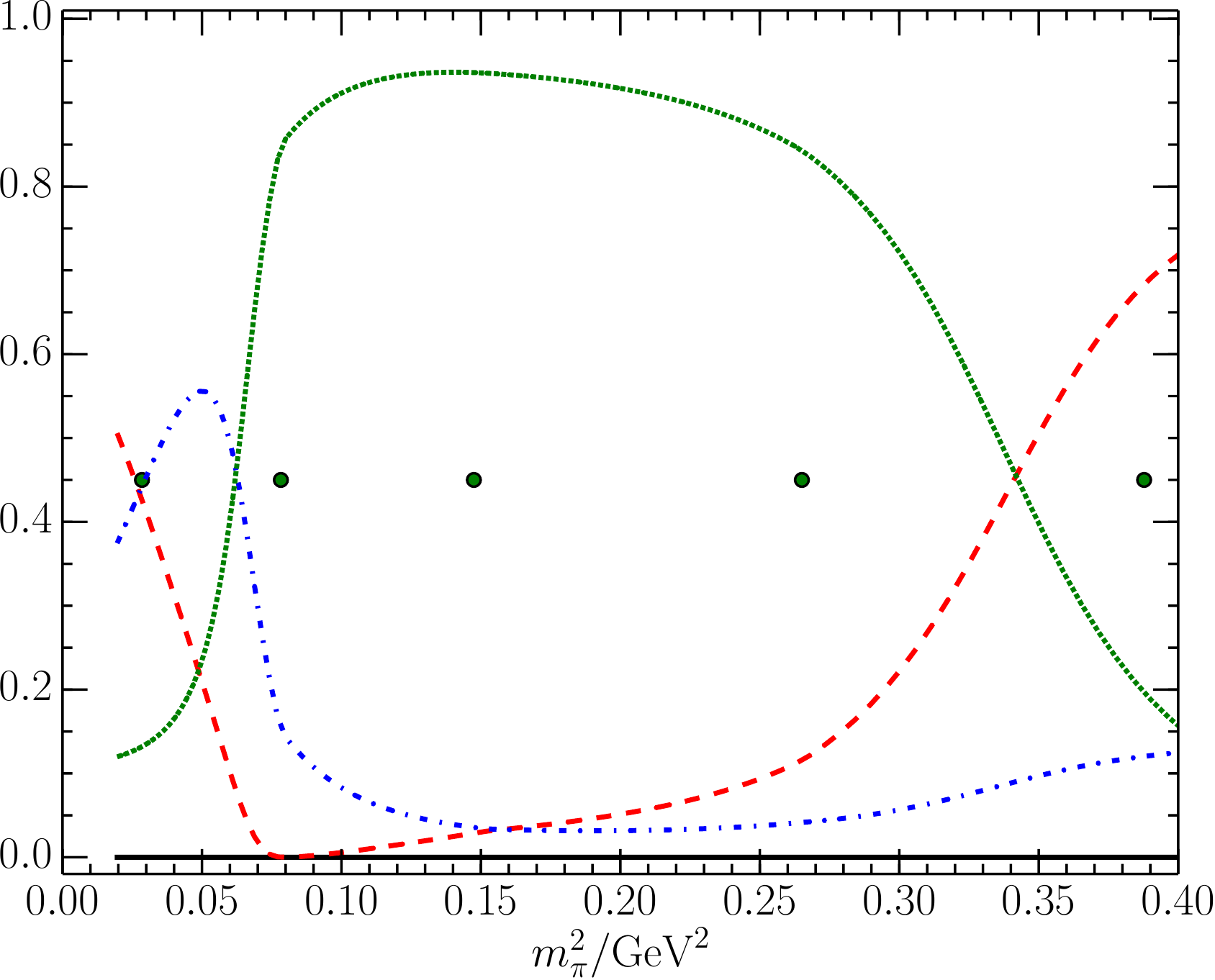}}\label{figComExp3fm8}}
\caption{{\bf Colour online:} The pion-mass evolution of the Hamiltonian eigenvector components for
  the first eight states observed in the scenario without a bare baryon state on the lattice volume
  with $L = 2.90$ fm.  The fifth and sixth states display a non-trivial superposition of all
  three basis states, $|\pi N\rangle$, $|\pi \Delta\rangle$, and $|\sigma N\rangle$ at light pion
  masses.}
\label{figCompnExp3fm}
\end{center} 
\end{figure}

The fractional meson-baryon components for the eight lowest-lying eigenstates of this scenario are
plotted in Fig.~\ref{figCompnExp3fm}.  Again, the lowest-lying state is predominantly $|\sigma
N\rangle$.  Noting that the second and third eigenstates are associated with the low-lying lattice
QCD results at large pion masses, Figs.~\ref{figComExp3fm2} and \ref{figComExp3fm3} indicate these
states are dominated by $|\pi N\rangle$ and $|\sigma N\rangle$ basis states.  After an avoided
level crossing at large pion masses, the composition of these two states is exchanged.

The fifth and sixth states of Figs.~\ref{figComExp3fm5} and \ref{figComExp3fm6} are more
interesting.  At light pion masses these states are a non-trivial superposition of all three basis
states, $|\pi N\rangle$, $|\pi \Delta\rangle$, and $|\sigma N\rangle$.  These states appear to be
more than weakly mixed scattering states and it is interesting that these are the levels consistent
with the lattice QCD results at light pion masses.
  
While only a few of the eigenstates illustrated in Fig.~\ref{figSpecExp3fm} have been seen on the
lattice, one should in principle be able to observe all of these states in future lattice QCD
calculations.  The key is to move beyond local three quark operators.  Five quark operators
\cite{Kiratidis2015} have successfully revealed low-lying scattering states in the odd-parity
nucleon channel that were missed with three-quark operators.  Moreover, five-quark multi-particle
operators where the momentum of both the meson and the baryon are projected at the source are
particularly efficient at exciting the lowest-lying scattering states \cite{Lang:2012db}.  Future
lattice QCD simulations will draw on these techniques to fill in the missing states predicted by
our Hamiltonian model.

In summary, scenario II describes the experimental scattering data and also describes the lattice
QCD results as non-trivial mixings of the basis states.  More trivial mixings of the basis states
are not seen on the large-volume lattice simulations because the overlap of weakly mixed two-particle
scattering states with local three-quark operators is suppressed by the spatial volume of the lattice.

\subsection{System with a Bare Nucleon Basis State} \label{secSBN}

In light of the success of our second scenario describing the Roper resonance as a pure molecular
meson-baryon state, we proceed to a third scenario in which these channels have an opportunity to
mix with the bare baryon state associated with the ground state nucleon.  There is {\it a priori}
no reason to omit such couplings.

We find fits to the the phase shifts and inelasticities to be rather insensitive to the couplings and mass of
the bare nucleon state, $|N_0\rangle$.  Thus, to constrain the couplings and the bare mass, we fit
the CSSM lattice QCD results simultaneously with the experimental phase shifts and inelasticities.
We restrict the cutoffs $\Lambda_\alpha$ of the exponential regulators to be the same as those in
the second scenario. In addition, we restrict the nucleon pole to be $939$ MeV at infinite volume.
We plot the best fit results for the  scattering data as dashed lines in Fig.~\ref{figPSEta}
and summarise the parameters in the right-hand column of Table \ref{tabPara} labeled III.  

To obtain the eigen-energy spectrum at finite volume, we need the nucleon mass as a function of the
squared pion mass, $m_N(m_\pi^2)$.  In the previous two scenarios, we used a linear interpolation between
the nucleon lattice results from the CSSM.  Here, we obtain $m_N(m_\pi^2)$ via iteration where the
lowest eigen-energy of the Hamiltonian model output is used as the input for $m_N(m_\pi^2)$ in the
next iteration.  Convergence is obtained without difficulty.
The slope of the bare nucleon mass as a function of $m_\pi^2$ is found to be
\begin{eqnarray}
\alpha_N^0=0.995~{\rm GeV}^{-1} \, .
\label{eqParaN0}
\end{eqnarray}

\begin{figure}[t]
	\begin{center}
		\includegraphics[width=1.0\columnwidth]{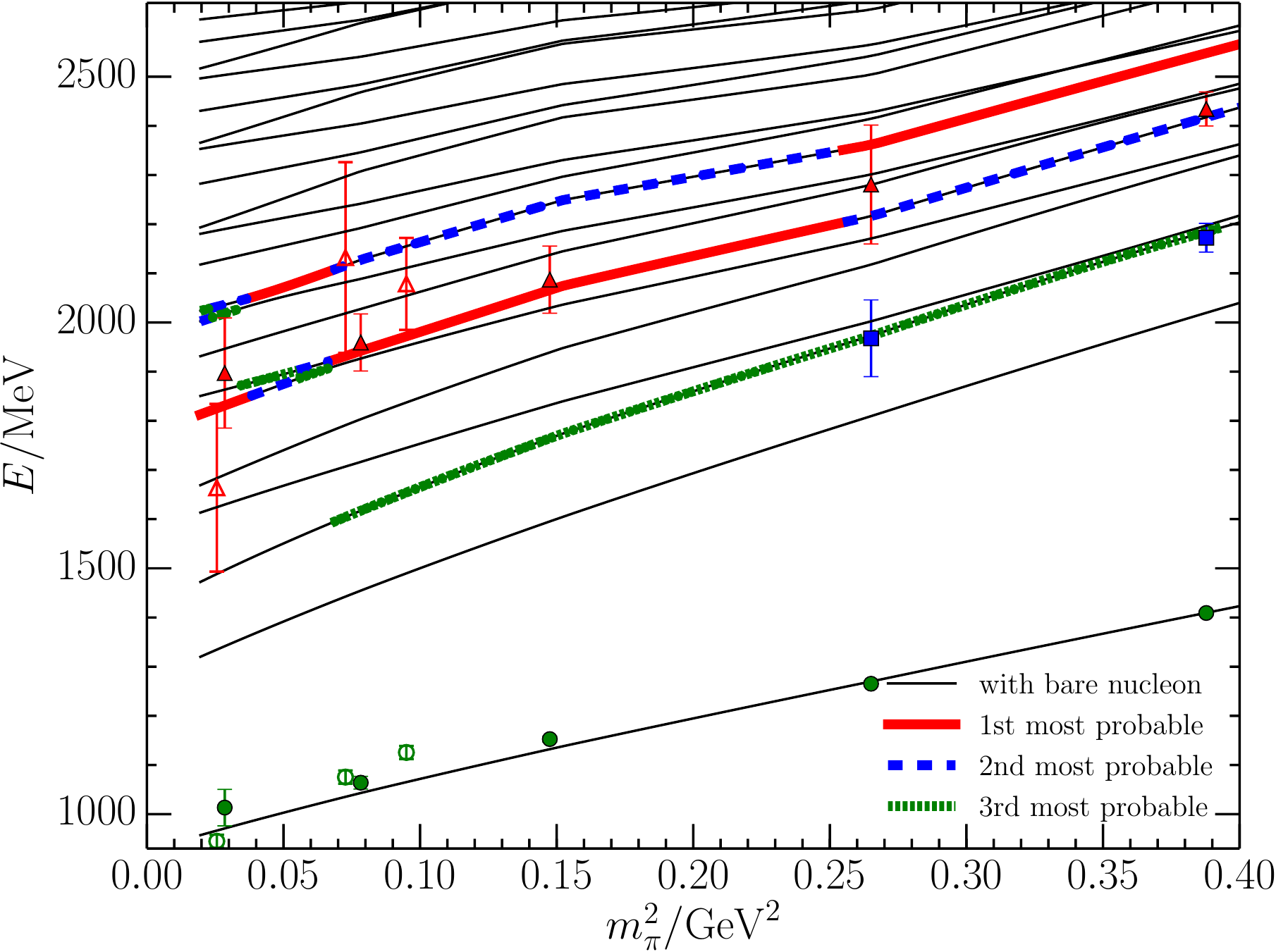}
		\caption{{\bf Colour online}: The pion mass dependence of the $L = 2.90$ fm finite-volume
			energy eigenstates for the Hamiltonian model scenario with a bare nucleon basis state.  As in
			Fig.~\ref{figSpecDpl3fm}, the different line types and colours used in illustrating the energy
			levels indicate the strength of the bare basis state in the Hamiltonian-model eigenvector.  The
			thick-solid (red), dashed (blue) and dotted (green) lines correspond to the states having the
			first, second, and third largest bare-state contributions and therefore represent the most
			probable states to be observed in the lattice QCD simulations.}
		\label{figSpecExpBN3fm}
	\end{center} 
\end{figure}

\begin{figure}[t]
	\begin{center}
		\includegraphics[width=1.0\columnwidth]{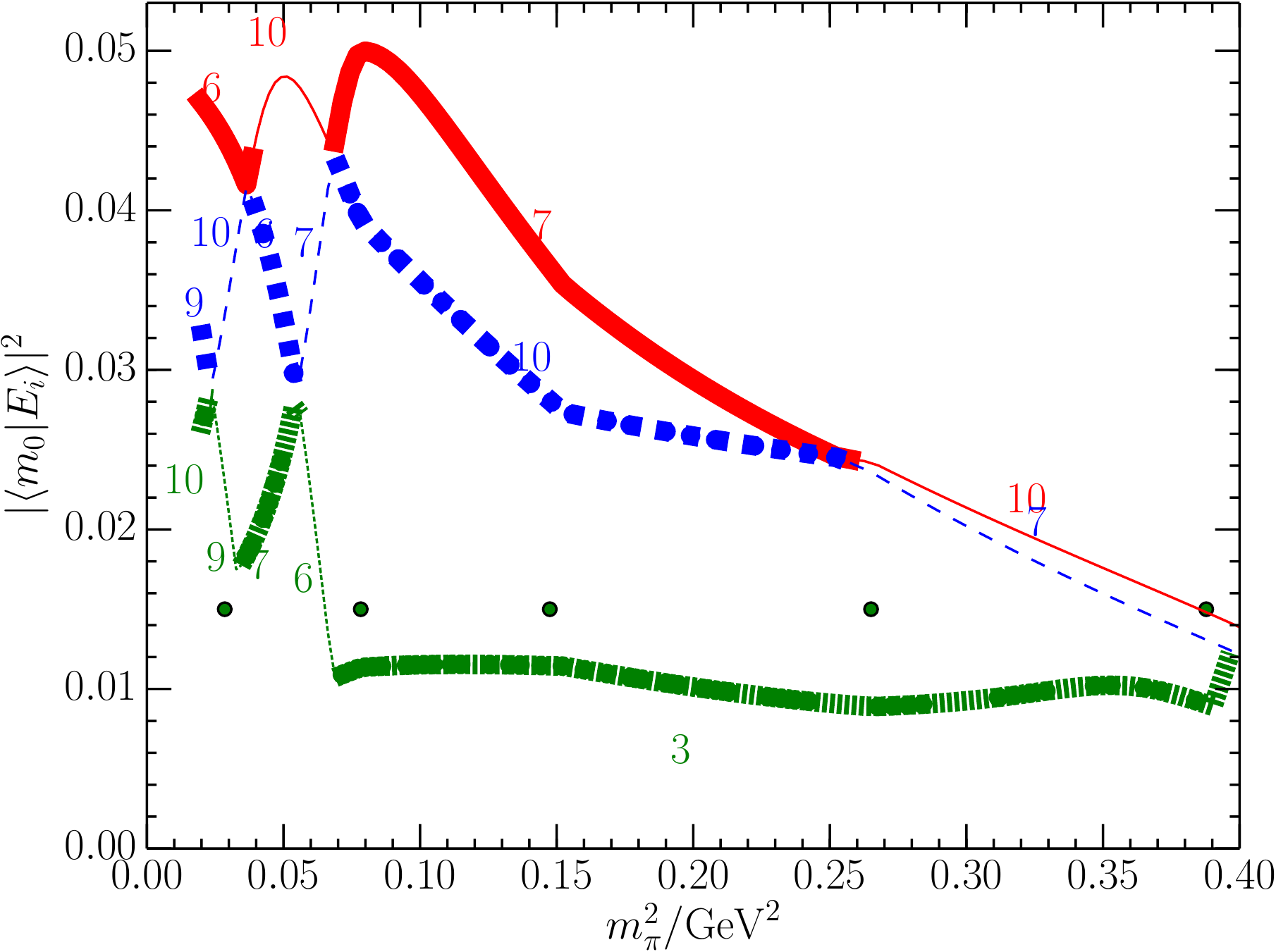}
		\caption{{\bf Colour online:} The fraction of the bare nucleon basis state, $| m_0 \rangle$, in the
			Hamiltonian energy eigenstates $| E_i \rangle$ for the three states having the largest bare-state
			contribution.  As in Fig.~\ref{figBareRDpl3fm}, states are labeled by the energy-eigenstate
			integers, $i$.  The dark-green dots plotted at $y = 0.015$ indicate the
			positions of the five quark masses considered in the CSSM lattice results.  While the line type
			and colour scheme matches that of Fig.~\ref{figSpecExpBN3fm}, the thick and thin lines alternate
			to indicate a change in the energy eigenstate associated with that value.}
		\label{figBareRExp3fm}
	\end{center} 
\end{figure}

The energy levels predicted by the Hamiltonian Model and the proportion of the bare nucleon basis
state in the excited states are illustrated in Figs.~\ref{figSpecExpBN3fm} and \ref{figBareRExp3fm}
respectively.
The CSSM ground-state nucleon data are fit well by the Hamiltonian model and those from the Cyprus
group are clustered near the ground-state curve in Fig.~\ref{figSpecExpBN3fm}.  We obtain the
nucleon mass on the $L \simeq 2.90$ fm volume at the physical pion mass of $0.140$ GeV to be
\begin{equation}
m_N(m_\pi^2|_{\rm phy})_{L\simeq 2.90~{\rm fm}}=0.957~{\rm GeV} \, , \quad
\end{equation}
revealing that the finite volume of the lattice increases the nucleon mass by nearly 20 MeV.
The nucleon ground state on the 3 fm lattice contains 80\% $\sim$ 90\% of the bare nucleon basis
state.

As in the first scenario, we anticipate that excited states having a large bare-state component will have a
more significant coupling with the three-quark operators used to excite the states in contemporary
lattice QCD calculations.  Figure \ref{figBareRExp3fm} identifies excited states having the largest
bare-state components and thus the most probable states to be seen on the lattice.

For example, at the lightest quark mass, the sixth energy eigenstate has the largest
bare-nucleon component and is the most likely state to be observed in current lattice QCD
calculations.  Correspondingly, the sixth excitation energy in Fig.~\ref{figSpecExpBN3fm} is
highlighted with a thick red line.
Both the CSSM and Cyprus lattice calculations produce an excited state consistent with the sixth
energy level.
Remarkably the second most probable state to be seen in lattice QCD simulations lies even higher in
energy at approximately 2 GeV.

For $m_\pi^2$ lying in the range $0.07 \sim 0.27\ {\rm GeV}^2$, the seventh eigenstate is predicted
to be the most easily seen with the proportion of bare nucleon basis state at 2.5\% $\sim$ 5\%.
The second most probable state to be seen in this regime is state 10 at 2.1 to 2.3 GeV.  For
$m_\pi^2 > 0.27\ {\rm GeV}^2$ states 7 and 10 contain roughly equal amounts of bare state
contributions.  The lattice QCD results are consistent with the lower-lying state of these two most
probable states to be seen in lattice QCD.
We do not analyse the lattice QCD results near the tenth eigen-energy level as this
energy regime surpasses the realm of our model constraints.

While the lattice QCD simulations reveal a low-lying state at the largest two quark masses
considered (blue filled squares in Fig.~\ref{figSpecExpBN3fm}) the trend does not continue into the
lighter quark mass regime.  Figure \ref{figBareRExp3fm} provides an explanation for this
observation.  For the second and third lightest quark masses at $m_\pi^2 \simeq 0.08$ and 0.15
GeV${}^2$ the bare-state contribution to the low-lying (green curve) state is three to five times
smaller than that for the states having the largest contribution.  This reduced bare-state
component is expected to coincide with a reduced coupling of the state to three-quark operators.
Thus a possible explanation for the omission of the lower-lying state at light quark masses in
current lattice QCD simulations is that its relatively small coupling to three-quark interpolators
is insufficient for it to be seen in the correlation matrix analysis with current levels of
statistical accuracy.

\begin{figure}[t]
\begin{center}
\subfigure[\ 1st eigenstate]{\scalebox{\sca}{\includegraphics{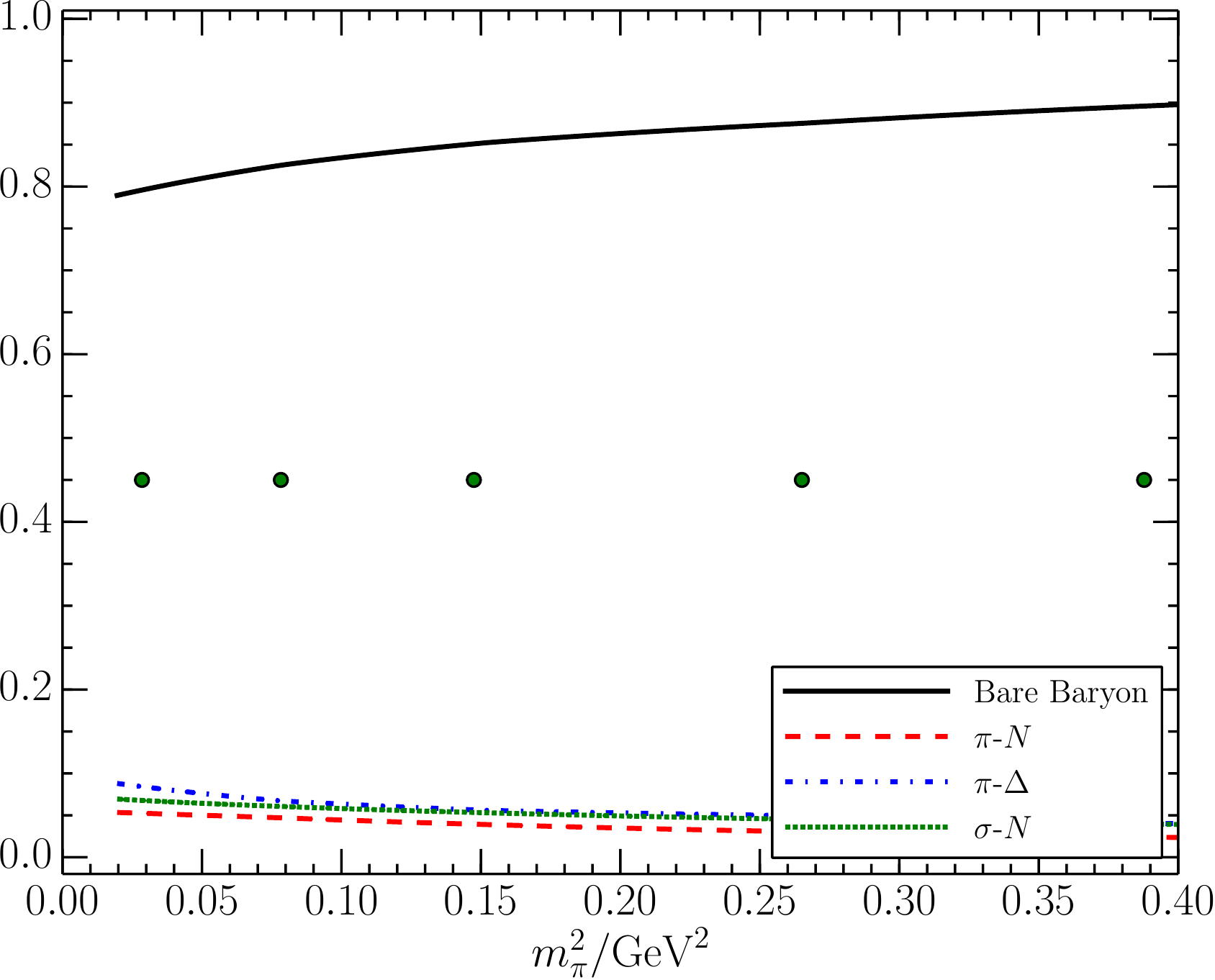}}\label{figCnExpBN1}}
\subfigure[\ 2nd eigenstate]{\scalebox{\sca}{\includegraphics{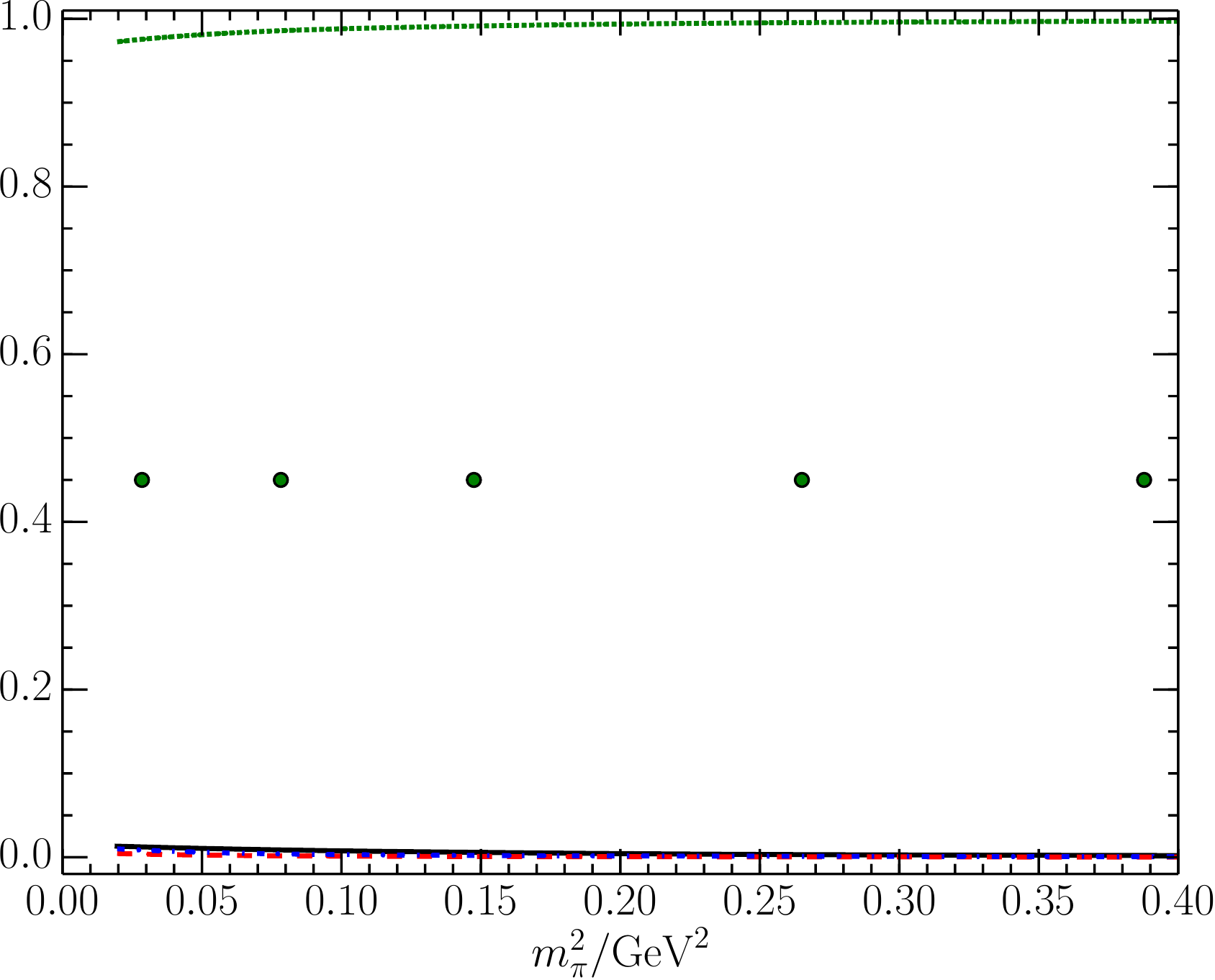}}\label{figCnExpBN2}}
\subfigure[\ 3rd eigenstate]{\scalebox{\sca}{\includegraphics{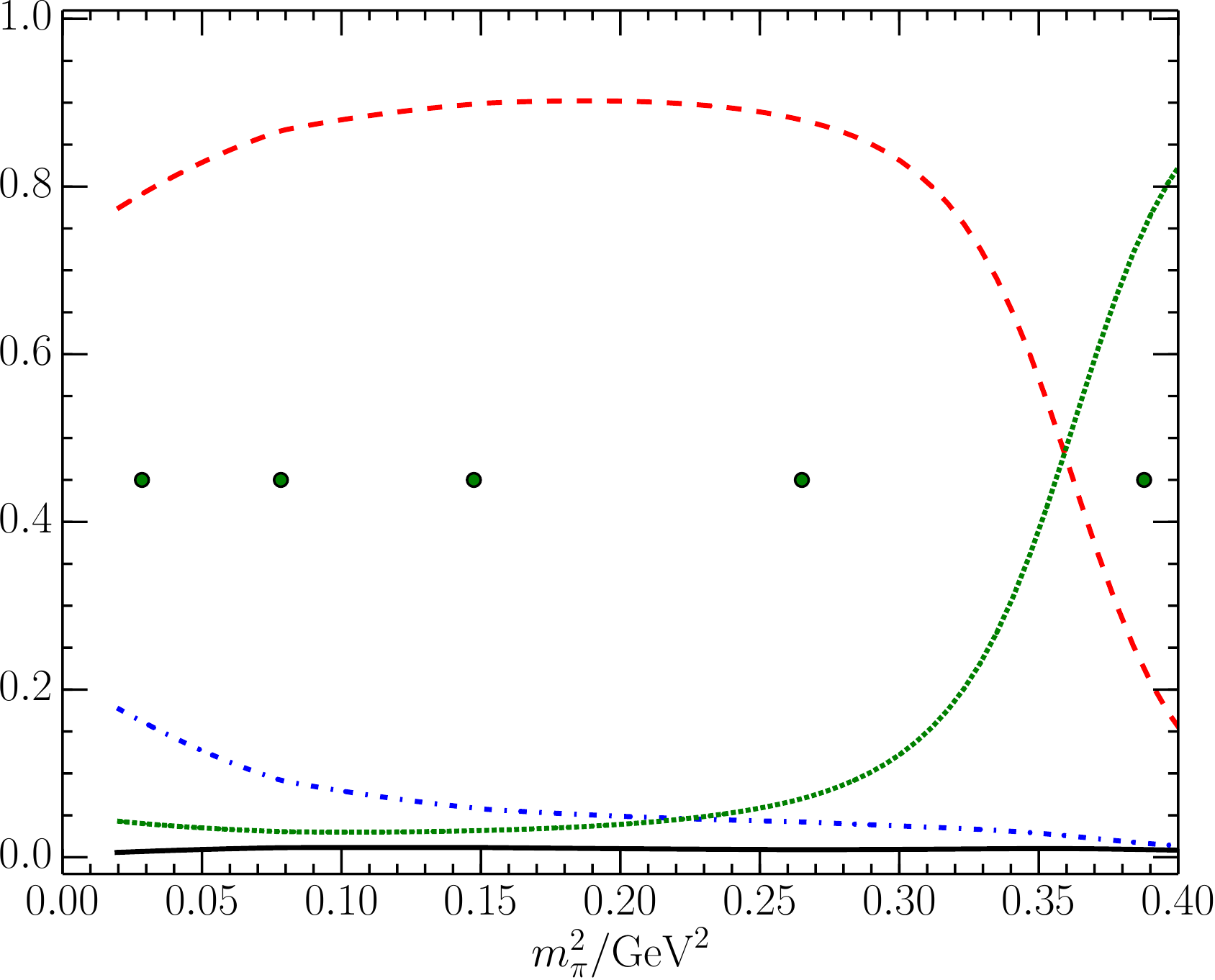}}\label{figCnExpBN3}}
\subfigure[\ 4th eigenstate]{\scalebox{\sca}{\includegraphics{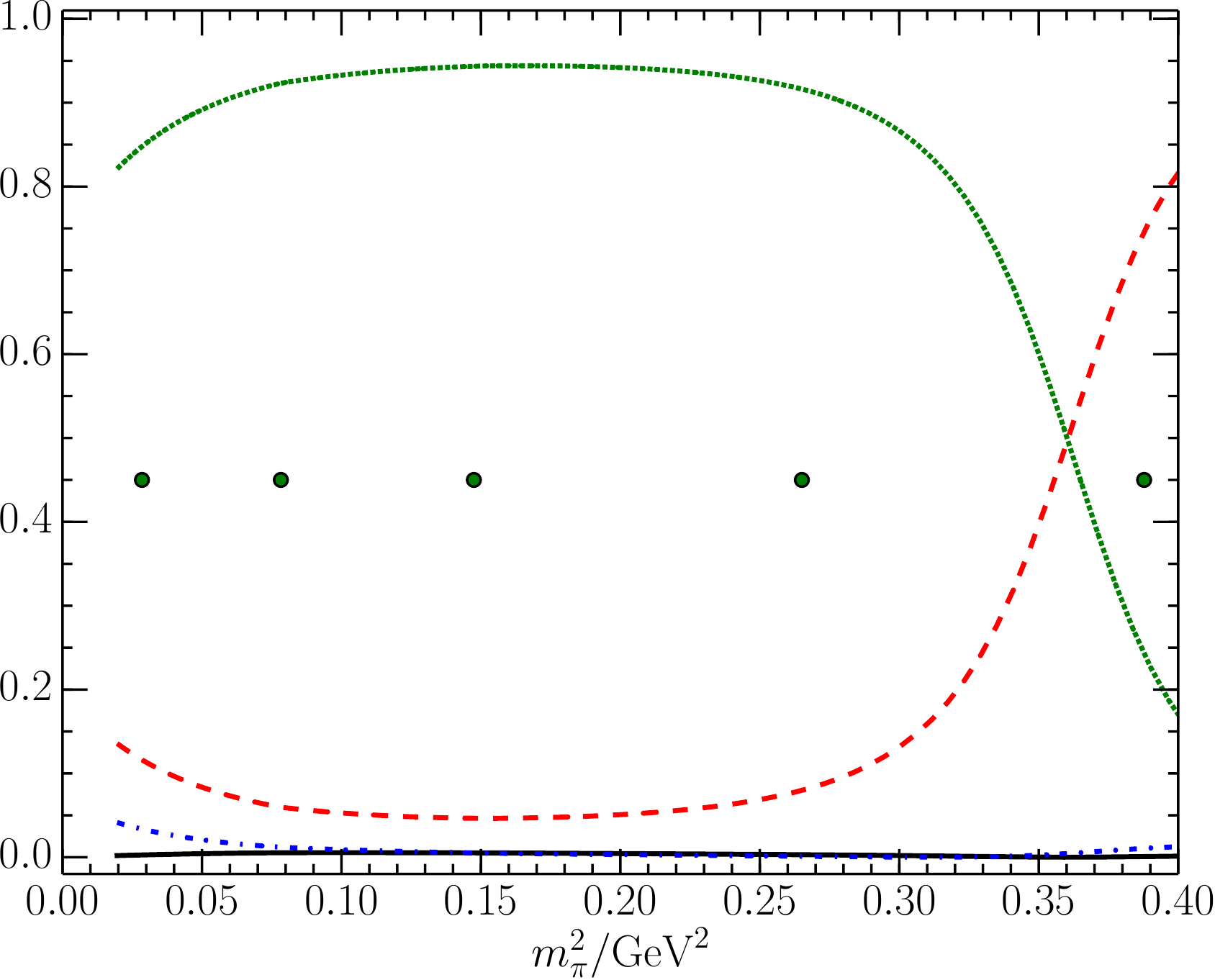}}\label{figCnExpBN4}}
\subfigure[\ 5th eigenstate]{\scalebox{\sca}{\includegraphics{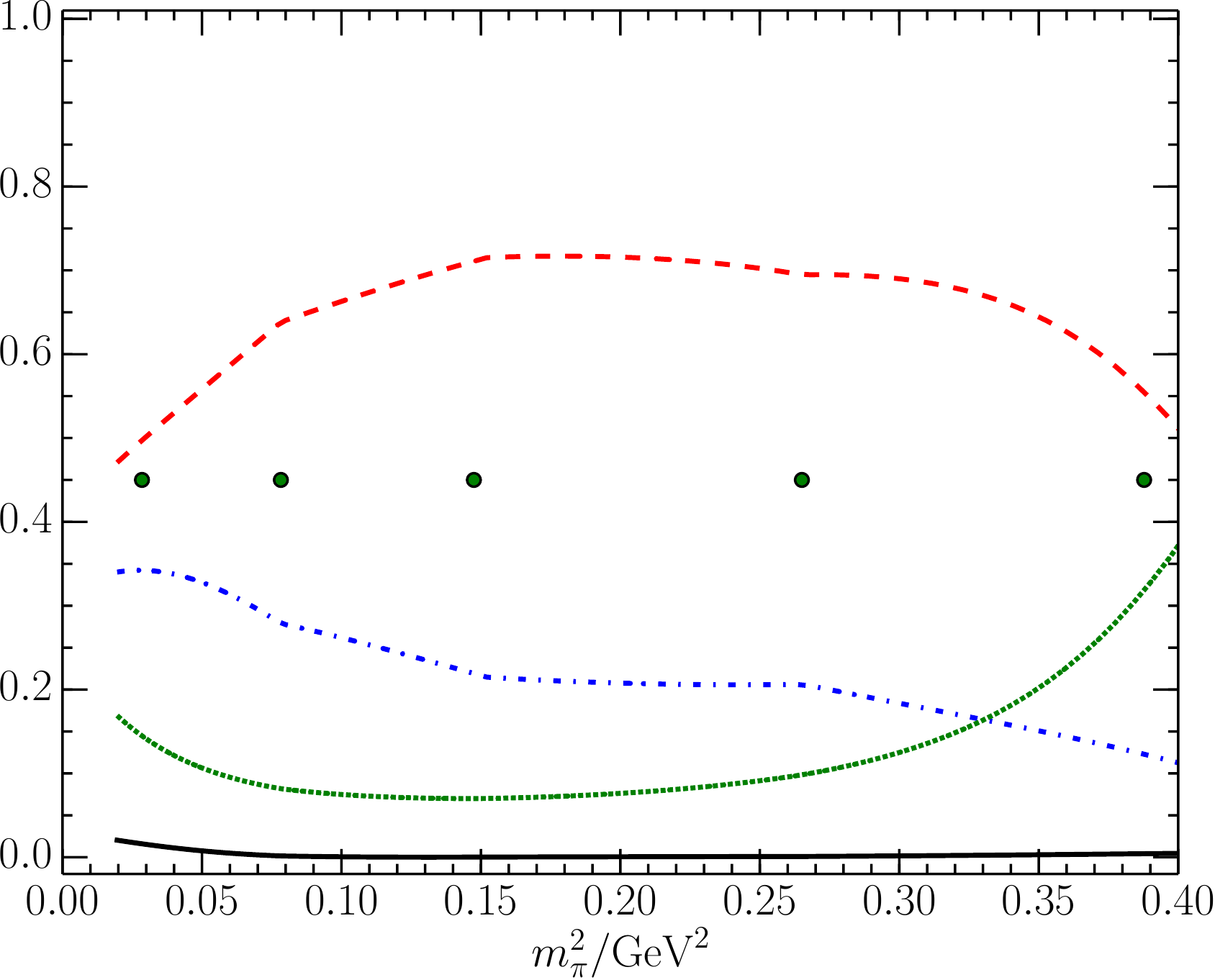}}\label{figCnExpBN5}}
\subfigure[\ 6th eigenstate]{\scalebox{\sca}{\includegraphics{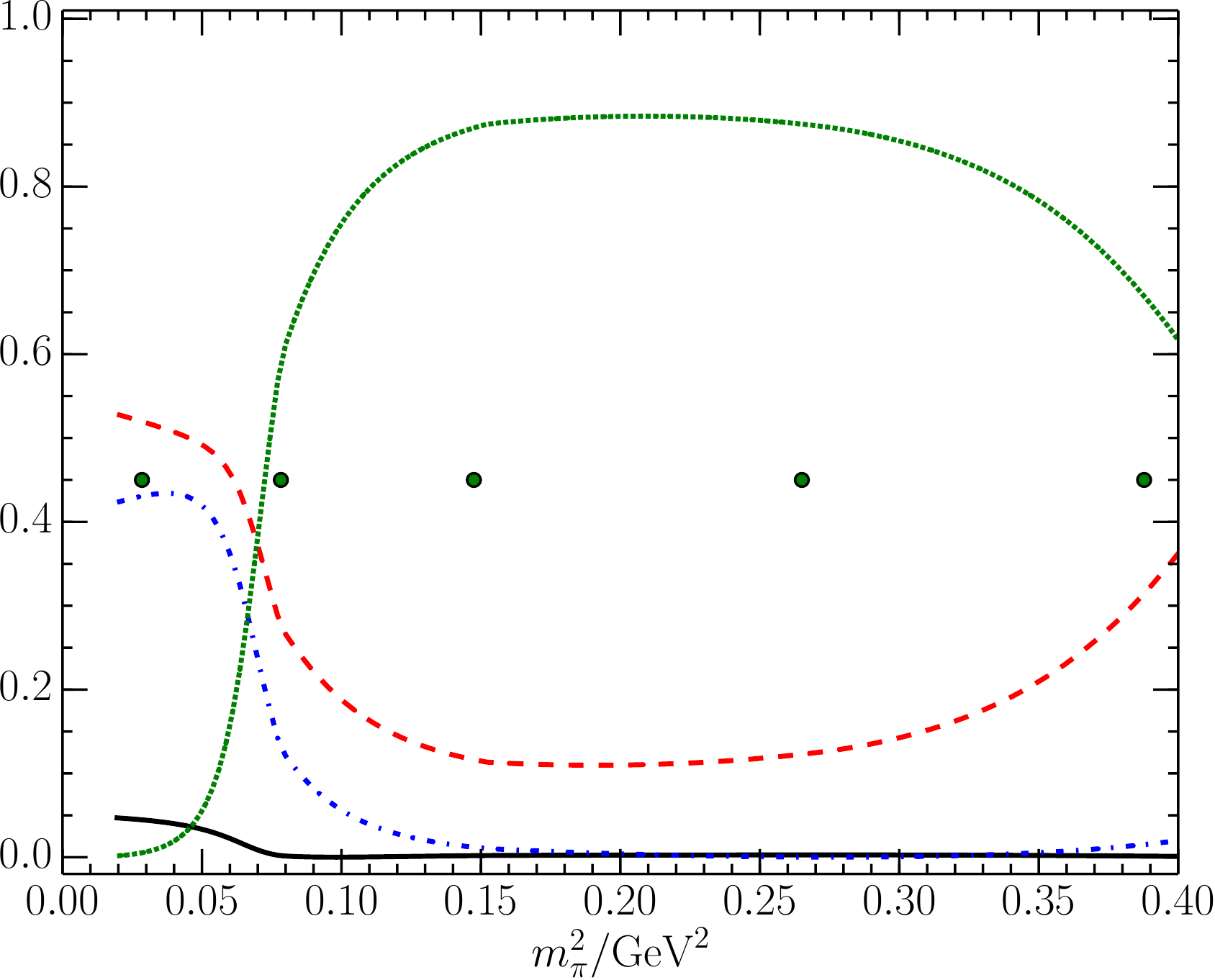}}\label{figCnExpBN6}}
\subfigure[\ 7th eigenstate]{\scalebox{\sca}{\includegraphics{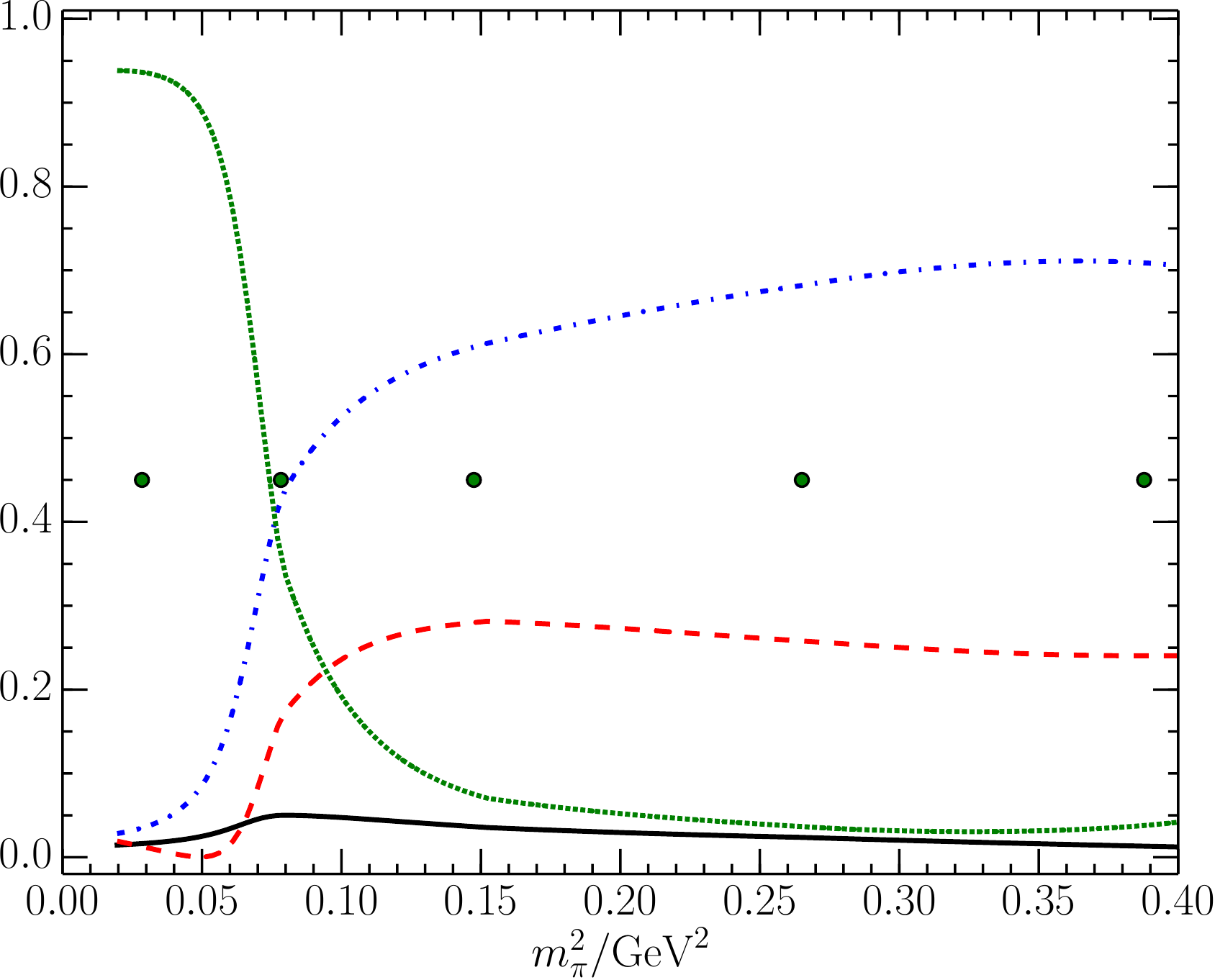}}\label{figCnExpBN7}}
\subfigure[\ 8th eigenstate]{\scalebox{\sca}{\includegraphics{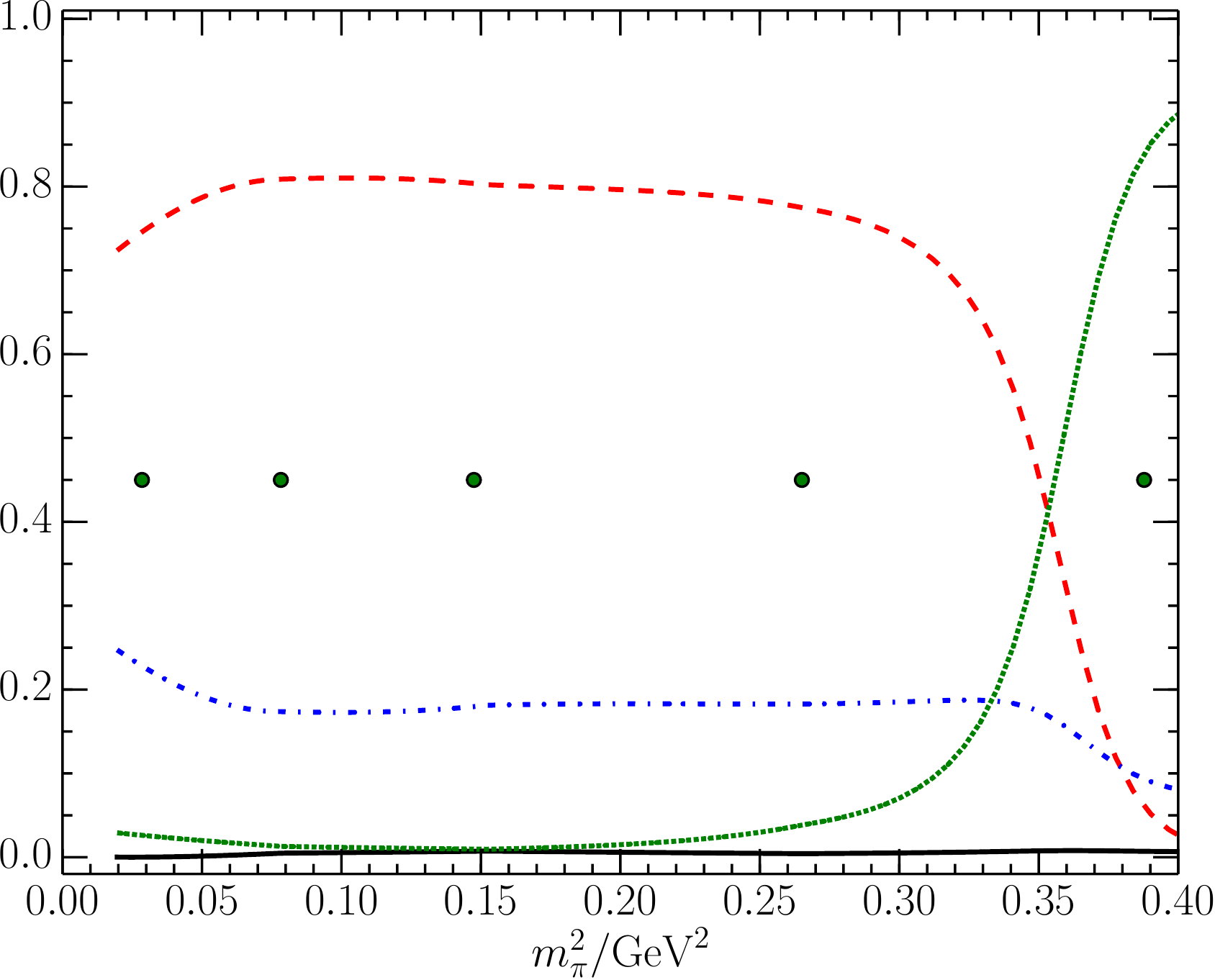}}\label{figCnExpBN8}}
\caption{{\bf Colour online:} The pion-mass evolution of the Hamiltonian eigenvector components for
  the first eight states (including the ground state) observed in the scenario with a bare nucleon
  basis state on the lattice volume with $L = 2.90$ fm.  Of all the excited states, the sixth
  and seventh states have the largest bare-state component at light quark masses.  This component
  is accompanied by a non-trivial superposition of meson-baryon basis states.}
\label{figCompnExpBN3fm}
\end{center} 
\end{figure}

The components of the eigenstate vectors are illustrated for this scenario in
Fig.~\ref{figCompnExpBN3fm}. 
At the physical pion mass, Fig.~\ref{figCnExpBN1} indicates the ground state nucleon is 80\% bare
nucleon dressed with 20\% meson-baryon states spread evenly over the three meson-baryon channels
considered.

As this analysis now includes the ground-state nucleon as the first state, the state labels have
changed in scenario III being one larger than in scenarios I and II.  
Once again the first excitation is predominantly $|\sigma N\rangle$.  

Noting that the second and third excitations (states 3 and 4) are in the realm of the low-lying
lattice QCD results at large pion masses, Figs.~\ref{figCnExpBN3} and \ref{figCnExpBN4} indicate
these states are dominated by $|\pi N\rangle$ and $|\sigma N\rangle$ basis states.  As in scenario
II an avoided level crossing at large pion masses causes the composition of these two states to be
exchanged.  However, state three has the largest bare state component at the two largest quark
masses considered and it is this state that has most likely been produced in the lattice
simulations.

The next excitation, state 5, resembles state 4 of scenario II being predominantly $|\pi N\rangle$.

The sixth and seventh eigenstates (corresponding to the fifth and sixth eigenstates in scenario II)
continue to show a non-trivial mixing of the meson-baryon basis states near the lightest two quark
masses considered by the CSSM.  It is precisely in this region of non-trivial mixing that the bare
state component becomes manifest.  
At the lightest quark mass, Fig.~\ref{figCnExpBN6} illustrates that the sixth eigenstate is 50\%
$|\pi N\rangle$, 45\% $|\pi \Delta\rangle$ and 5\% bare nucleon.
At the second lightest quark mass considered by the CSSM, the seventh state in Fig.~\ref{figCnExpBN7} has the largest
bare-state component with 
40\% $|\pi \Delta\rangle$, 
40\% $|\sigma N \rangle$,
15\% $|\pi N\rangle$
and 
5\% bare nucleon.
It is precisely these excited states having the largest bare state component that correspond to the
states seen in lattice QCD simulations at light quark masses.

\subsection{Comparison of the Three Scenarios}

We have studied the phase shifts and inelasticities at infinite volume and the finite-volume
eigenstates on the lattice in three scenarios, a bare Roper basis state in the first, no bare
baryon basis state in the second and a bare nucleon basis state in the third scenario.  As
illustrated in Fig.~\ref{figPSEta}, there are differences in the phase shifts and inelasticities
among the three scenarios.  Within the constraints of these models, it is not possible to find
sets of fit parameters which can make the phase shifts and inelasticities of the three scenarios
overlap everywhere.  Moreover, the fits cannot give the same phase shifts and inelasticities in each
of the $\pi N$, $\pi\Delta$, and $\sigma N$ channels.  This will directly lead to different
eigen-energy spectra on the lattice, based on L\"uscher's Theorem.

\begin{figure}[t]
\begin{center}
\includegraphics[width=1.0\columnwidth]{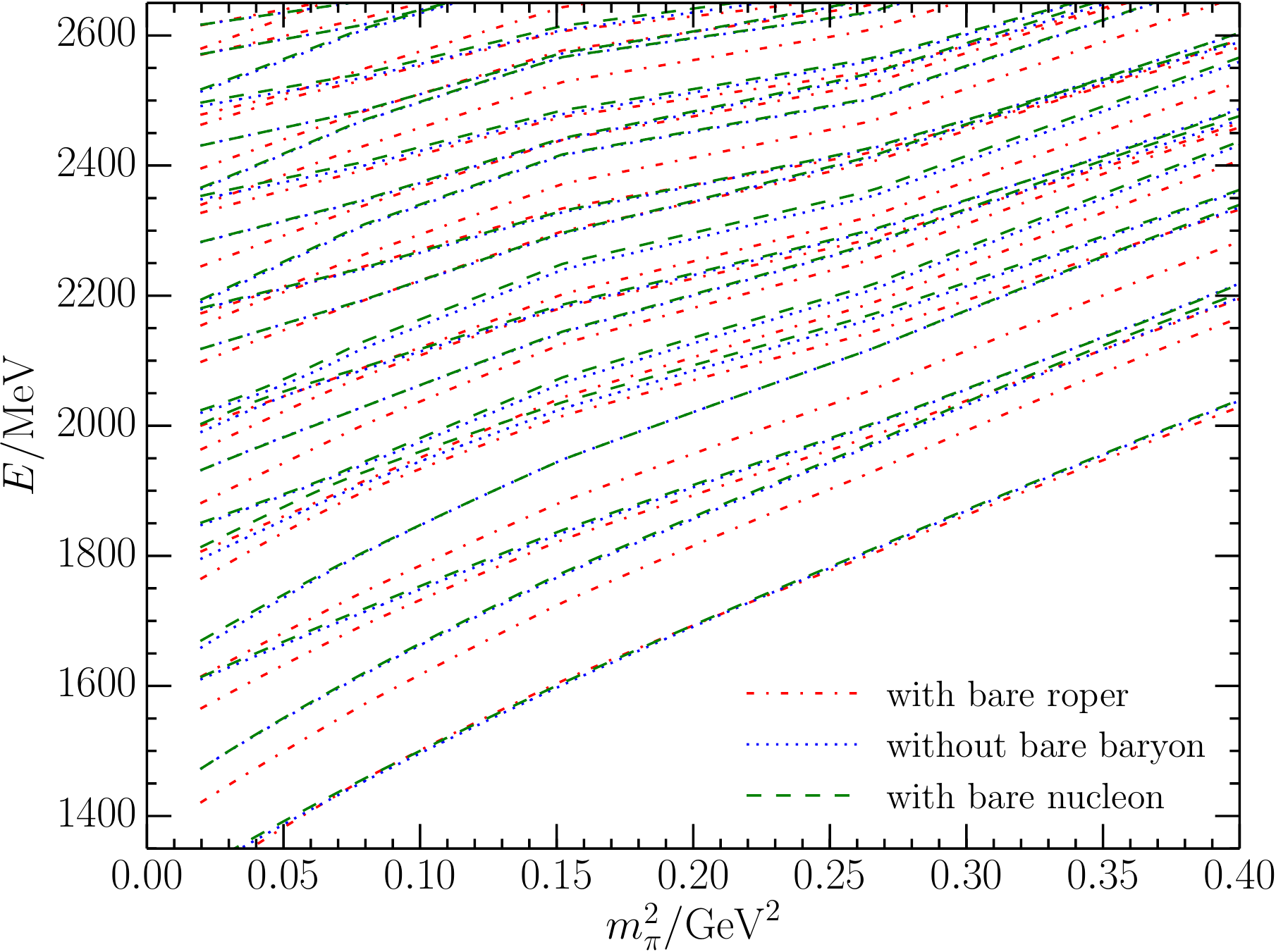}
\caption{{\bf Colour online:} Comparison of the excited-state energy levels for the three scenarios
  at finite volume with $L = 2.90$ fm. All scenarios use $m_N(m_\pi^2)$ obtained in the third
  scenario.}
\label{figCompare3fm}
\end{center} 
\end{figure}

In performing a direct comparison of the energy levels predicted in our three scenarios, the pion
mass dependence of the ground state nucleon mass $m_N(m_\pi^2)$ obtained in scenario III is used
in all three scenarios.  The energy levels for the $L = 2.90$ fm lattice are compared in
Fig.~\ref{figCompare3fm}.

We can see a significant difference between cases with and without the bare Roper basis state in
Fig. \ref{figCompare3fm}.  However, differences between scenarios III and II, with and without a
bare nucleon basis state respectively, are subtle.  The main feature provided through the inclusion
of a bare nucleon basis state is a clear understanding of the states to be observed in contemporary
lattice QCD calculations.

\subsection{Results for a Volume with $\mathbf{L\simeq 1.98}$ fm}

In this section, we consider the smaller spatial lattice volume with $L \simeq 1.98$ fm considered
by the Hadron Spectrum Collaboration (HSC) \cite{Edwards2011}.  Drawing on the fit results to the
experimental data summarized in Table \ref{tabPara}, one can proceed to explore the predictions of
the Hamiltonian model on this very small volume lattice.

\begin{figure}[t]
\begin{center}
\subfigure[\ Spectrum]
{\scalebox{\sca}{\includegraphics{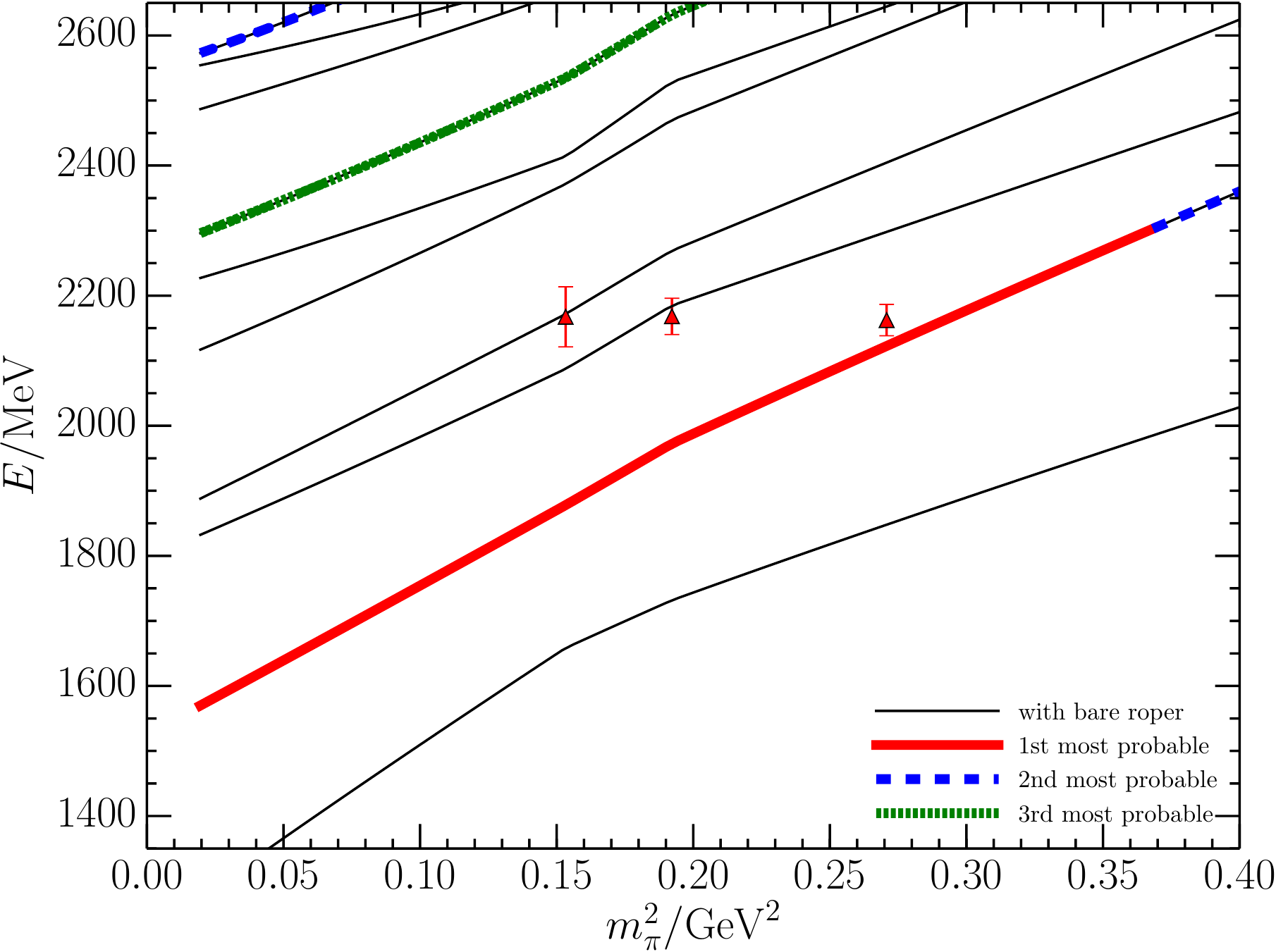}}\label{figSpec2fmR}}
\subfigure[\ Bare Roper Component]
{\scalebox{\sca}{\includegraphics{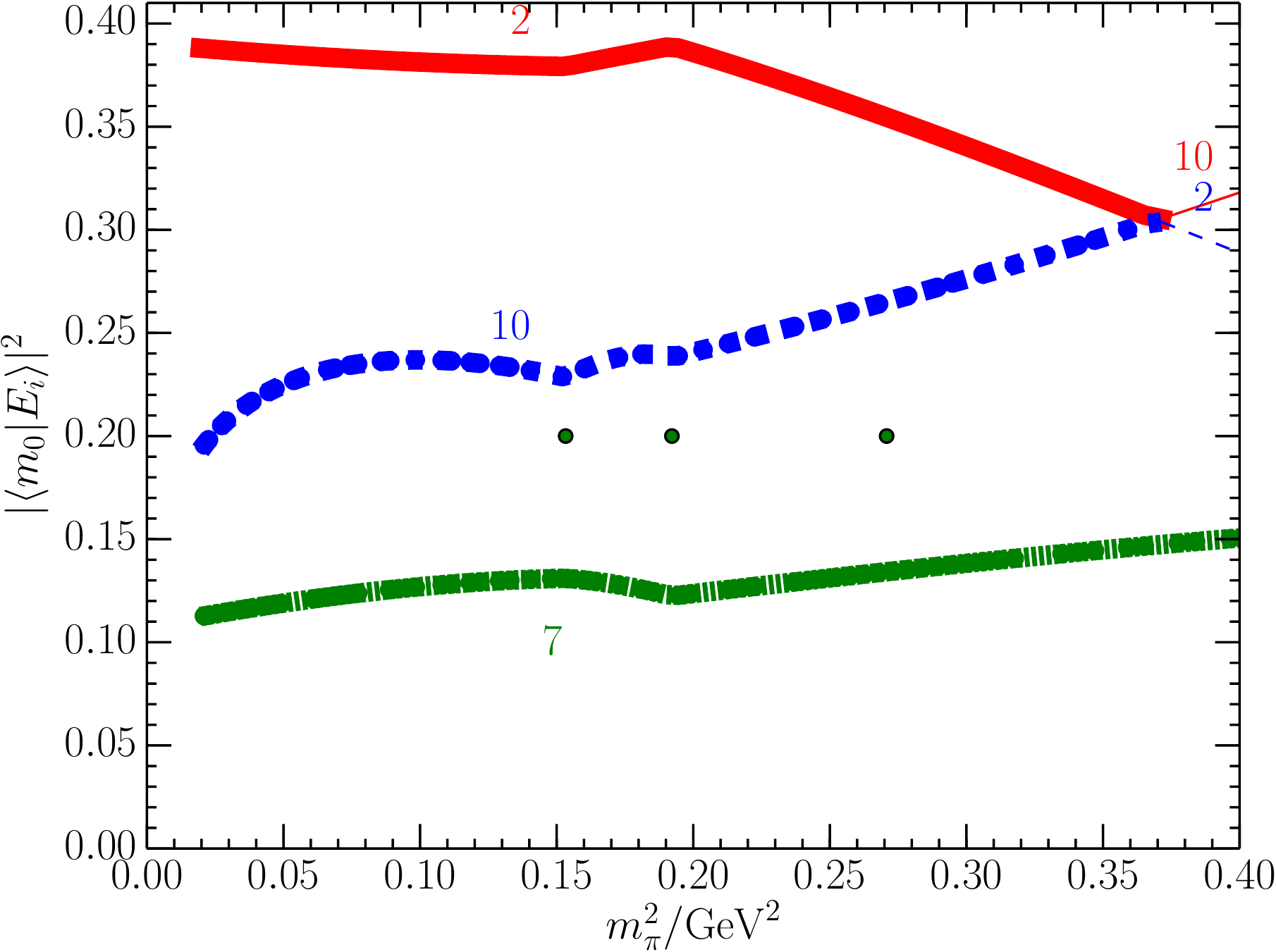}}\label{figBareR2fmR}}
\caption{{\bf Colour online:} The pion mass dependence of the $L = 1.98$ fm finite-volume energy
  eigenstates (left) and the fraction of the bare-Roper basis state, $| m_0 \rangle$, in the
  Hamiltonian energy eigenstates $| E_i \rangle$ for the three states having the largest bare-state
  contribution (right) for the Hamiltonian model scenario with a bare Roper basis state.  
  The spectrum results with filled symbols (left) are from the Hadron Spectrum Collaboration
  \cite{Edwards2011}.  
  The dark-green dots plotted at $y = 0.30$ (right) indicate the positions of the three quark
  masses considered in the HSC results.  
  The line types and colour schemes match those of Figs.~\ref{figSpecDpl3fm} and
  \ref{figBareRDpl3fm}.  
}
\end{center} 
\end{figure}

First we consider Scenario I with the bare Roper basis state.  The finite-volume spectrum of states
are compared with the HSC results in Fig.~\ref{figSpec2fmR} while the proportion of the bare Roper
basis state is illustrated in Fig.~\ref{figBareR2fmR}.  The HSC results sit near the energy levels
of the matrix Hamiltonian model.  However, the same problem encountered in the $L = 2.90$ fm volume
case appears here.  The Hamiltonian model predicts states approaching 1.6 GeV in the light
quark-mass regime having a bare state component exceeding 35\% of the eigenvector.  Such a state
should be easy to excite with the three-quark operators considered by the HSC.  The absence of such
a state at the two lightest quark masses considered by the HSC in the mass range 1.85 to 2.00 GeV
provides further evidence that the Roper resonance is not composed with a bare basis state with mass
$\simeq 2.0$ GeV.

\begin{figure}[htbp]
\begin{center}
\subfigure[\ Spectrum]
{\scalebox{\sca}{\includegraphics{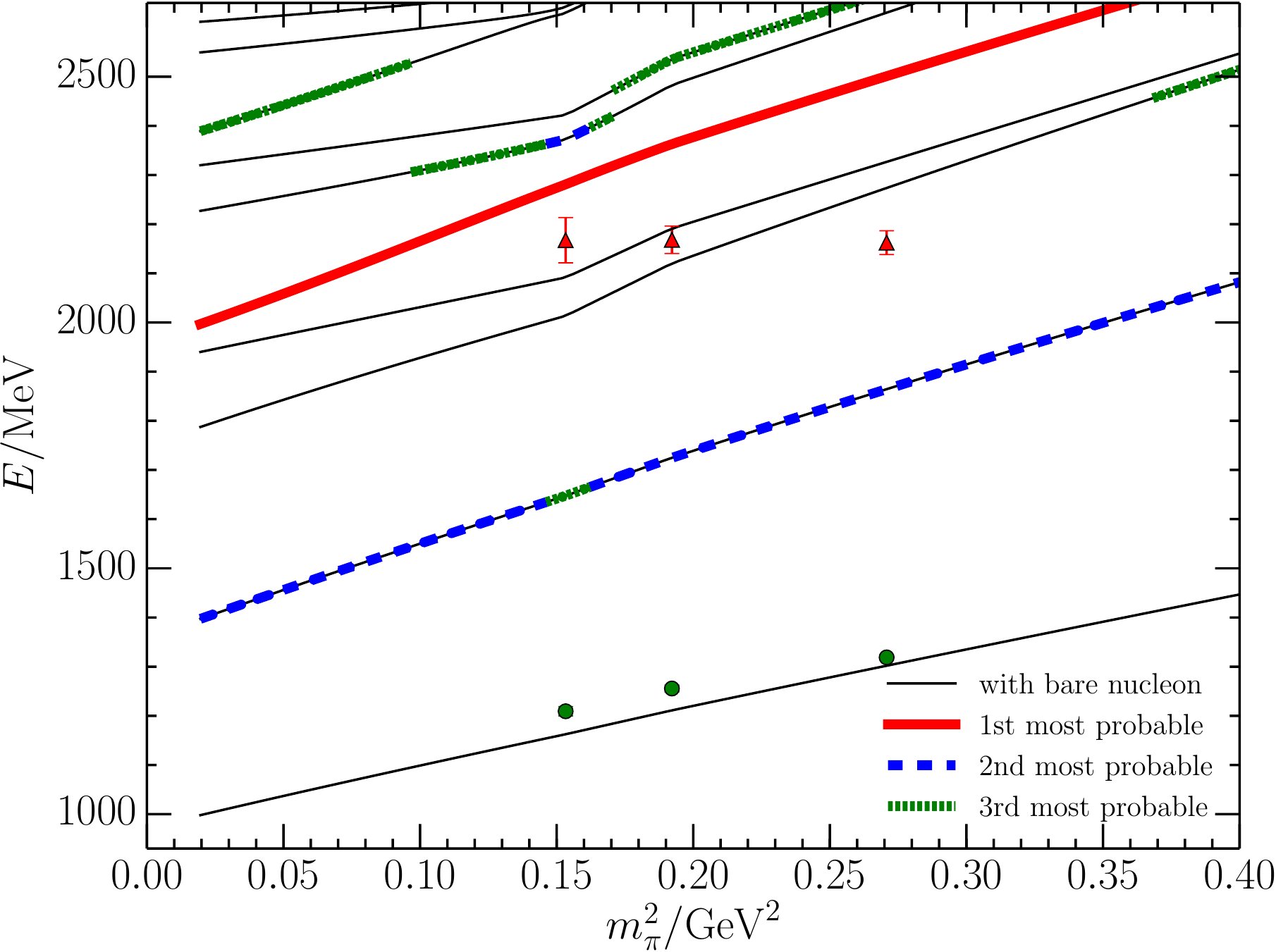}}\label{figSpec2fmN}}
\subfigure[\ Bare Nucleon Component]
{\scalebox{\sca}{\includegraphics{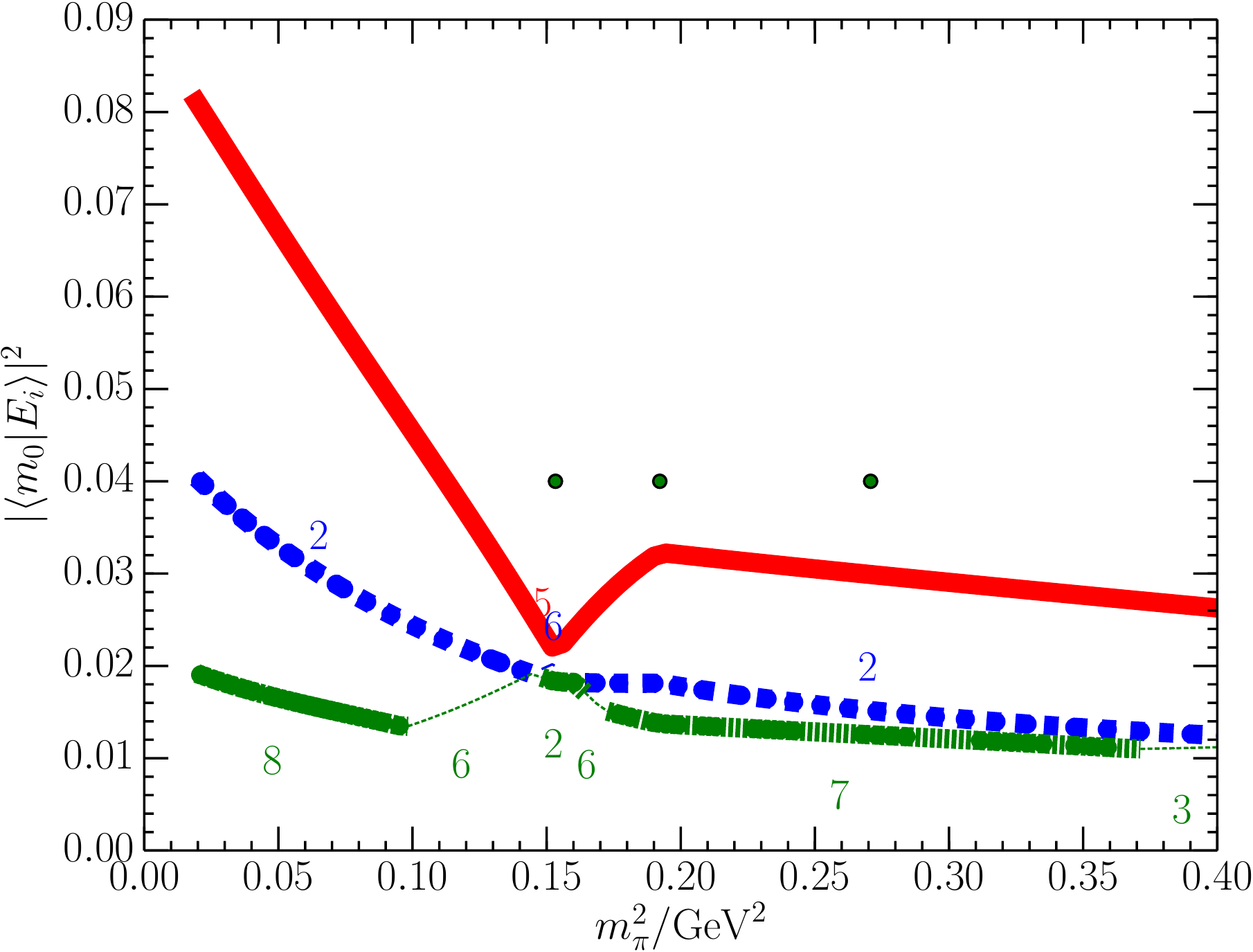}}\label{figBareR2fmN}}
\caption{{\bf Colour online:} 
  The pion mass dependence of the $L = 1.98$ fm finite-volume energy
  eigenstates (left) and the fraction of the bare-Roper basis state, $| m_0 \rangle$, in the
  Hamiltonian energy eigenstates $| E_i \rangle$ for the three states having the largest bare-state
  contribution (right) for the Hamiltonian model scenario with a bare Nucleon basis state.  
  The line types, symbols and colour schemes are as in Figs.~\ref{figSpec2fmR} and
  \ref{figBareR2fmR}.  
}
\label{fig2fmN}
\end{center} 
\end{figure}

As for the $L = 2.90$ fm results, there is little difference in the finite volume spectra of
Scenarios II and III, and therefore we proceed directly to an illustration of the results for
Scenario III.  The results with the bare nucleon basis state in the finite volume of $L\simeq 1.98$
fm are presented in Fig.~\ref{fig2fmN}.   This time the HSC results do not coincide with the 
finite volume energy levels of the Hamiltonian model illustrated in Fig.~\ref{figSpec2fmN}.  There
are two finite-volume based concerns that can contribute to the origin of this discrepancy.

One concern is the interference of the coarse infrared momentum discretisation induced by the small
periodic volume and the ultraviolet regulators of the loop integrals constrained by the
experimental phase shifts and inelasticities of Fig.~\ref{figPSEta}.  For $P$-wave meson-baryon
dressings, the zero-momentum contribution is absent and the finite volume acts as an infrared
regulator.  The form factors in the third scenario are constrained by experiment to have small
volume such that even the first momenta available on the small value are already suppressed, almost
rendering the meson-baryon dressings negligible.

For example, for a finite volume with $L=2$ fm and $k_i=2\pi/L$
\begin{equation}
u_{\pi N}^{\rm III}(k_1)=0.296,~ 
u_{\pi N}^{\rm III}(k_2)=0.088,~
u_{\pi N}^{\rm III}(k_3)=0.026,
\end{equation}
in scenario III.  These form factors are small compared to those of the first scenario on the
same volume
\begin{equation}
u_{\pi N}^{\rm I}(k_1)=0.456,~ 
u_{\pi N}^{\rm I}(k_2)=0.208,~
u_{\pi N}^{\rm I}(k_3)=0.095.
\end{equation}
They are also significantly smaller than those of the third scenario with $L=3$ fm where
\begin{equation}
u_{\pi N}^{\rm III}(k_1)=0.582,~ 
u_{\pi N}^{\rm III}(k_2)=0.339,~
u_{\pi N}^{\rm III}(k_3)=0.197.
\end{equation}

\begin{figure}[t]
\begin{center}
\includegraphics[clip=true,trim=2.9cm 0.0cm 2.9cm 0.0cm,width=0.48\columnwidth]{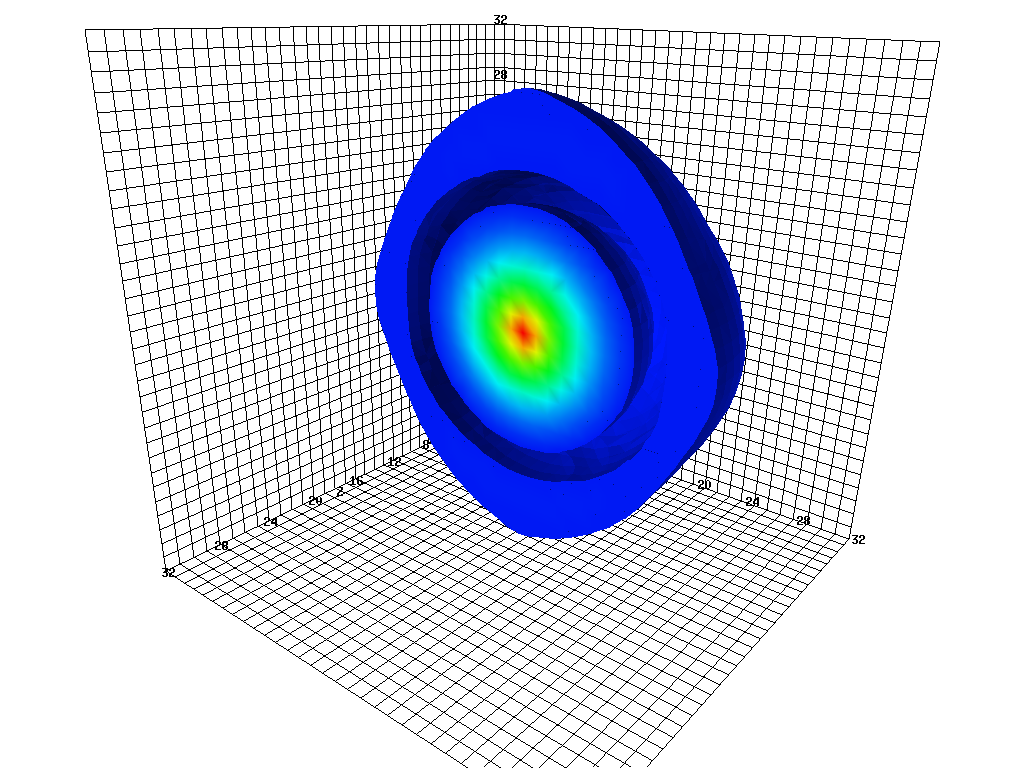}
\includegraphics[clip=true,trim=2.9cm 0.0cm 2.9cm 0.0cm,width=0.48\columnwidth]{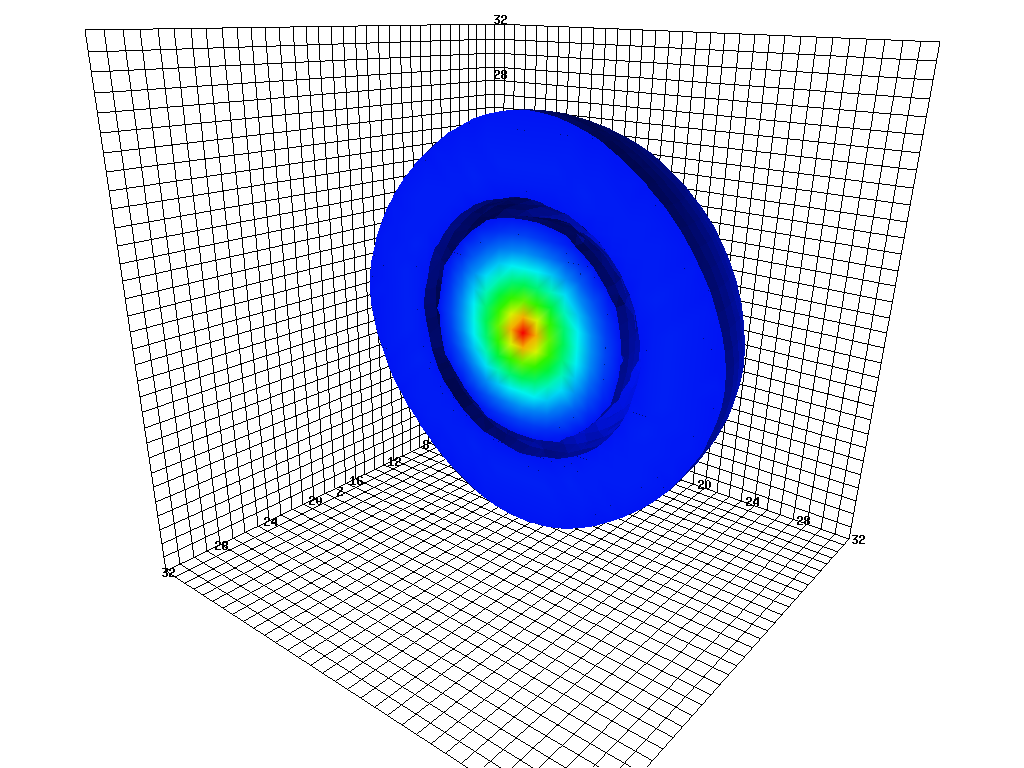}
\caption{{\bf Colour online:} Lattice QCD calculation of the $d$-quark probability distribution in
  the first excited state of the proton, reproduced from Ref.~\cite{Roberts:2013oea}.  The CSSM's
  lattice calculations at the (left) lightest quark mass (state 6 in scenario III) and (right)
  second heaviest quark mass (state 3 in scenario III) reveal the effect of the finite periodic
  volume of the lattice with $L \simeq 2.90$ fm.  Here, two $u$ quarks are fixed at the origin at the centre
  of the plot and the lattice spacing is approximately 0.09 fm. }
\label{figWavefunction}
\end{center} 
\end{figure}

Another concern is that the effects of the small finite volume may induce distortions that cannot
be accounted for by meson-baryon dressings alone.  Figure \ref{figWavefunction} reproduced from
Ref.~\cite{Roberts:2013oea} illustrates the influence of the periodic volume of the $d$-quark
probability distribution in the first excited state of the nucleon.  This lattice QCD calculation
was carried out on lattice volumes with $L \simeq 2.90$ fm.  Even at large quark masses
corresponding to $m_\pi^2 = 0.27$ GeV${}^2$ distortions in the spherically symmetric $2S$ radial
wave function are readily observed.  They become even more severe as the value of the isovolume cut
is lowered and the tails of the wave function are considered \cite{Roberts:2013oea}.  Therefore we
caution that an even smaller volume with $L \simeq 1.98$ fm is likely to have finite volume
distortions that cannot be described by effective field theory alone.

\section{Conclusion}\label{secSum}

We have studied the infinite volume phase shifts and inelasticities for scattering states with
$I(J^P)=\frac12(\frac12^+)$ in effective field theory.  Through the consideration of a
finite-volume Hamiltonian Matrix we have explored the corresponding finite-volume spectra of states
relevant to contemporary lattice QCD calculations of the spectrum.

In doing so we have explored three scenarios for the underlying theory describing the available
data.  All three scenarios are able to describe the scattering data well and all three create a
pole position for Roper similar to that reported by the PDG.  However the finite volume spectrum
predicted by the scenarios has important differences.

In the first case, the Roper is postulated to have a triquark-like bare or core component with a
mass exceeding the resonance mass.  This component mixes with attractive virtual meson-baryon
contributions, including the $\pi N$, $\pi \Delta$, and $\sigma N$ channels, to reproduce the
observed pole position.

With the advent of new insight from the Hamiltonian model and lattice QCD results we have been able to discard
this popular description of including a bare Roper basis state.  This model predicts a low-lying state
in the finite volume having a very large bare-state component that makes it accessible to current
lattice QCD techniques.  The absence of this state in today's lattice QCD calculations exposes an
inconsistency in the model predictions.

In the second hypothesis, the Roper resonance is dynamically generated purely from the $\pi N$,
$\pi \Delta$, and $\sigma N$ channels in the absence of a bare-baryon basis state.  This scenario
identifies the lattice QCD results as non-trivial superpositions of the basis states that have a
qualitative difference from the weak mixing of basis states in the scattering channels.  However,
given the presence of a bare state associated with the ground state nucleon, we proceed to consider
a third scenario incorporating the presence of this basis state.

In the third scenario the Roper resonance is composed of the low-lying bare basis component
associated with the ground state nucleon.  The merit of this scenario lies in its ability to not
only identify and describe the finite-volume energy levels to be observed in contemporary
large-volume lattice QCD simulations but also explain why other low-lying states have been missed
in today's lattice QCD results for the nucleon spectrum.

We conclude that the Roper resonance of Nature is predominantly a dynamically-generated molecular
meson-baryon state with a weak coupling to a low-lying bare basis state associated with the ground
state nucleon.  

This conclusion is in sharp contrast to a conventional state with a large three-quark core
component like the ground state nucleon or even the $N(1535)$ resonance where a significant
bare-state contribution was manifest \cite{Liu:2015ktc}.  It also suggests that relativistic
three-quark bound state approaches \cite{Eichmann:2016yit} will fail as these models do not have
the full influence of the meson-baryon sector required to generate the full coupled-channel
physics.

Future work should investigate the role of three-body coupled channel effects in the structure of
the Roper resonance.  Of particular interest in the role of the $N(\pi \pi)_\texttt{S-wave}$
channel.  While our consideration of the $\sigma N$ channel does model the effects of the $N(\pi
\pi)_\texttt{S-wave}$ channel, the $\sigma$ meson is a broad state with a large width and it is
desirable to accommodate this important physics in a more direct manner.  For example the imaginary
part of the Roper pole position is likely to be sensitive to this physics.

Similarly, it may be interesting to explore other models of the Roper resonance and their finite
volume implementation.  For example one could further explore the nature of the bare basis state
and its impact on resonance structure.

It is also desirable to advance lattice QCD simulations to include five-quark interpolating
fields where the momentum of each of the meson-baryon pairs can be defined at the source.  Not only
does this approach address the volume suppression of multi-particle states through a double sum in
the Fourier projection, it also enables the creation of a state very similar to the scattering
state in the finite volume of the lattice.  With this approach it should be possible to observe all
the states predicted by the Hamiltonian model and eventually reverse the process such that the
experimental phase shifts and inelasticities are determined from the finite volume spectra of
lattice QCD.
Such developments will be key in obtaining a full understanding of the Roper resonance.

After this paper was submitted, a very important lattice QCD simulation was released by Lang {\it
  et al.} \cite{Lang:2016hnn}.  In addition to standard three-quark operators, these authors included
explicit momentum-projected $\pi N$ and $\sigma N$ interpolating fields in a lattice QCD
analysis of the Roper channel.  The $\sigma N$ operator was included to simulate the effect of the
$N \pi \pi$ channel.  By comparing their energy levels with those calculated here they reached
similar conclusions.  Their results provide strong support for the third scenario and disfavor the
first scenario considered herein. The success of the Hamiltonian effective field theory in predicting
the position of these energy levels confirms the consideration of resonant two-body channels (such as the $\sigma N$ and $\pi \Delta$ channels) is effective in linking lattice QCD results to the Roper resonance of nature.  We note the inclusion of the $\pi \Delta$ contributions is essential for describing the inelasticity of the $\pi N$ to $\pi N$ amplitude.  We anticipate that when the $\pi \Delta$ channel is explored in future lattice QCD simulations, a new low-lying energy level will be observed consistent with our 5'th state at 1.7 GeV for the lightest quark mass.

\begin{acknowledgments}
We would like to thank T.-S.H. Lee for helpful discussions.  
We thank the PACS-CS Collaboration for making the $2+1$ flavor configurations underpinning the
current investigation available and the ILDG for their ongoing support in the sharing of these
configurations.
This research was undertaken with the assistance of resources at the NCI National Facility in
Canberra, the iVEC facilities at the Pawsey Centre and the Phoenix GPU cluster at the University of
Adelaide, Australia.  These resources were provided through the National Computational Merit
Allocation Scheme, supported by the Australian Government, and the University of Adelaide through
their support of the NCI Partner Share and the Phoenix GPU cluster.
This research is supported by the Australian Research Council through the ARC Centre of Excellence
for Particle Physics at the Terascale (CE110001104), and through Grants No.\ LE160100051, DP151103101 (A.W.T.), DP150103164,
DP120104627 and LE120100181 (D.B.L.).
\end{acknowledgments}


\begin{thebibliography}{10}
	
	\bibitem{Roper1964}
	L.~D. Roper,  Phys. Rev. Lett. \textbf{12}, 340 (1964).
	
	\bibitem{Aznauryan2008}
	I.~G. Aznauryan, \emph{et~al.} (CLAS Collaboration),  Phys. Rev. C \textbf{78},
	045209 (2008).
	
	\bibitem{Joo2005}
	K.~Joo, \emph{et~al.} (CLAS Collaboration),  Phys. Rev. C \textbf{72}, 058202
	(2005).
	
	\bibitem{PDG2014}
	K.~Olive and et~al. (Particle Data Group),  Chin.Phys.C \textbf{38}, 090001
	(2014).
	
	\bibitem{Mokeev2012}
	V.~I. Mokeev, \emph{et~al.} (CLAS Collaboration),  Phys. Rev. C \textbf{86},
	035203 (2012).
	
	\bibitem{Aznauryan2012}
	I.~Aznauryan and V.~Burkert,  Progress in Particle and Nuclear Physics
	\textbf{67}, 1  (2012).
	
	\bibitem{Thiel2015}
	A.~Thiel, \emph{et~al.} ((CBELSA/TAPS Collaboration)),  Phys. Rev. Lett.
	\textbf{114}, 091803 (2015).
	
	\bibitem{Weber1990}
	H.~J. Weber,  Phys. Rev. C \textbf{41}, 2783 (1990).
	
	\bibitem{Julia-Diaz2006}
	B.~Juli{\'a}-D{\'i}az and D.~Riska,  Nuclear Physics A \textbf{780}, 175
	(2006).
	
	\bibitem{Barquilla-Cano2007}
	D.~Barquilla-Cano, A.~J. Buchmann, and E.~Hern\'andez,  Phys. Rev. C
	\textbf{75}, 065203 (2007).
	
	\bibitem{Golli:2007sa} 
	B.~Golli and S.~Sirca, Eur. Phys. J. A \textbf{38}, 271 (2008).
	
	
	\bibitem{Golli:2009uk} 
	B.~Golli, S.~Sirca and M.~Fiolhais, Eur. Phys. J. A {\bf 42}, 185 (2009)
	
	
	\bibitem{Meissner1984}
	U.-G. Meissner and J.~Durso,  Nuclear Physics A \textbf{430}, 670  (1984).
	
	\bibitem{Hajduk1984}
	C.~Hajduk and B.~Schwesinger,  Physics Letters B \textbf{140}, 172  (1984).
	
	\bibitem{Krehl2000}
	O.~Krehl, C.~Hanhart, S.~Krewald, and J.~Speth,  Phys. Rev. C \textbf{62},
	025207 (2000).
	
	\bibitem{Schuetz1998}
	C.~Sch\"utz, J.~Haidenbauer, J.~Speth, and J.~W. Durso,  Phys. Rev. C
	\textbf{57}, 1464 (1998).
	
	\bibitem{Matsuyama2007}
	A.~Matsuyama, T.~Sato, and T.-S. Lee,  Physics Reports \textbf{439}, 193
	(2007).
	
	\bibitem{Kamano2010}
	H.~Kamano, S.~X. Nakamura, T.-S.~H. Lee, and T.~Sato,  Phys. Rev. C
	\textbf{81}, 065207 (2010).
	
	\bibitem{Kamano2013}
	H.~Kamano, S.~X. Nakamura, T.-S.~H. Lee, and T.~Sato,  Phys. Rev. C
	\textbf{88}, 035209 (2013).
	
	\bibitem{Hernandez2002a}
	E.~Hern\'andez, E.~Oset, and M.~J. Vicente~Vacas,  Phys. Rev. C \textbf{66},
	065201 (2002).
	
	\bibitem{Barnes1983}
	T.~Barnes and F.~Close,  Physics Letters B \textbf{123}, 89  (1983).
	
	\bibitem{Golowich1983}
	E.~Golowich, E.~Haqq, and G.~Karl,  Phys. Rev. D \textbf{28}, 160 (1983); [{\bf
		33}, 859 (1986)].
	
	\bibitem{Kisslinger1995}
	L.~S. Kisslinger and Z.~Li,  Phys. Rev. D \textbf{51}, R5986 (1995).
	
	\bibitem{Sarantsev2008}
	A.~Sarantsev, \emph{et~al.},  Physics Letters B \textbf{659}, 94  (2008).
	
	\bibitem{Mahbub:2013ala}
	M.~S. Mahbub, \emph{et~al.},  Phys. Rev. \textbf{D87}, 094506 (2013).
	
	\bibitem{Mahbub:2010rm}
	M.~S. Mahbub, \emph{et~al.} (CSSM Lattice),  Phys. Lett. \textbf{B707}, 389
	(2012).
	
	\bibitem{Roberts:2013ipa}
	D.~S. Roberts, W.~Kamleh, and D.~B. Leinweber,  Phys. Lett. \textbf{B725}, 164
	(2013).
	
	\bibitem{Roberts:2013oea}
	D.~S. Roberts, W.~Kamleh, and D.~B. Leinweber,  Phys. Rev. \textbf{D89}, 074501
	(2014).
	
	\bibitem{Alexandrou2015}
	C.~Alexandrou, T.~Leontiou, N.~Papanicolas, C.\, and E.~Stiliaris,  Phys. Rev.
	D \textbf{91}, 014506 (2015).
	
	\bibitem{Edwards2011}
	R.~G. Edwards, J.~J. Dudek, D.~G. Richards, and S.~J. Wallace,  Phys. Rev. D
	\textbf{84}, 074508 (2011).
	
	\bibitem{Kiratidis2015}
	A.~L. Kiratidis, W.~Kamleh, D.~B. Leinweber, and B.~J. Owen,  Phys. Rev. D
	\textbf{91}, 094509 (2015).
	
	\bibitem{Liu2014a}
	K.-F. Liu, \emph{et~al.},  PoS \textbf{LATTICE2013}, 507 (2014).
	
	\bibitem{Hall2013}
	J.~M.~M. Hall, \emph{et~al.},  Phys. Rev. D \textbf{87}, 094510 (2013).
	
	\bibitem{Wu2014}
	J.-J. Wu, T.-S.~H. Lee, A.~W. Thomas, and R.~D. Young,  Phys. Rev. C
	\textbf{90}, 055206 (2014).
	
	\bibitem{Luscher1990}
	M.~L{\"u}scher and U.~Wolff,  Nucl. Phys. B \textbf{339}, 222 (1990).
	
	\bibitem{Luescher1991}
	M.~L\"uscher,  Nuclear Physics B \textbf{354}, 531  (1991).
	
	\bibitem{Thomas1984}
	A.~W. Thomas,  Adv. Nucl. Phys. \textbf{13}, 1 (1984).
	
	\bibitem{Leinweber:2003dg}
	D.~B. Leinweber, A.~W. Thomas, and R.~D. Young,  Phys. Rev. Lett. \textbf{92},
	242002 (2004).
	
	\bibitem{Wang:2007iw}
	P.~Wang, D.~B. Leinweber, A.~W. Thomas, and R.~D. Young,  Phys. Rev.
	\textbf{D75}, 073012 (2007).
	
	\bibitem{Hall2015}
	M.~Hall, Jonathan~M.\, \emph{et~al.},  Phys. Rev. Lett. \textbf{114}, 132002
	(2015).
	
	\bibitem{Cloet:2002eg}
	I.~C. Cloet, D.~B. Leinweber, and A.~W. Thomas,  Phys. Rev. \textbf{C65},
	062201 (2002).
	
	\bibitem{Leinweber:2015kyz}
	D.~Leinweber, \emph{et~al.},  JPS Conf. Proc. \textbf{10}, 010011 (2016).
	
	\bibitem{Aoki:2008sm}
	S.~Aoki \emph{et~al.} (PACS-CS),  Phys. Rev. D \textbf{79}, 034503 (2009).
	
	\bibitem{Lang:2012db}
	C.~Lang and V.~Verduci,  Phys.Rev. \textbf{D87}, 054502 (2013).
	
	\bibitem{Liu:2015ktc}
	Z.-W. Liu, \emph{et~al.},  Phys. Rev. Lett. \textbf{116}, 082004 (2016).
	
	\bibitem{Eichmann:2016yit}
	G.~Eichmann, \emph{et~al.}, arXiv: 1606.09602  (2016).
	
	\bibitem{Lang:2016hnn}
	C. B. Lang, L. Leskovec, M. Padmanath, and S. Prelovsek, arXiv: 1610.01422 (2016).
\end{thebibliography}
\end{document}